\documentclass[twocolumn]{autart}    % Enable this line and disable the 
\usepackage{amsfonts,graphicx,amsmath,enumerate,color,amssymb,tabularx,subfigure,bm,booktabs}
\usepackage[longnamesfirst]{natbib}	
\usepackage[mathscr]{euscript}			 

\newcommand{\CCBB}{\mathbb{C}}

\newcommand{\ccSF}{\mathsf{c}}
\newcommand{\ddSF}{\mathsf{d}}

\newcommand{\iiSF}{\mathsf{i}}

\newcommand{\mmSF}{\mathsf{m}}
\newcommand{\nnSF}{\mathsf{n}}

\newcommand{\ppSF}{\mathsf{p}}
\newcommand{\qqSF}{\mathsf{q}}

\newcommand{\wwSF}{\mathsf{w}}

\newcommand{\zzSF}{\mathsf{z}}

\newcommand{\KKSF}{\mathsf{K}}

\newcommand{\TTCAL}{\mathcal{T}}

\newcommand{\ZZCAL}{\mathcal{Z}}

\newcommand{\BBRM}{\mathrm{B}}
\newcommand{\CCRM}{\mathrm{C}}
\newcommand{\DDRM}{\mathrm{D}}

\newcommand{\LLRM}{\mathrm{L}}
\newcommand{\MMRM}{\mathrm{M}}
\newcommand{\NNRM}{\mathrm{N}}

\newcommand{\PPRM}{\mathrm{P}}
\newcommand{\QQRM}{\mathrm{Q}}

\newcommand{\VVRM}{\mathrm{V}}
\newcommand{\WWRM}{\mathrm{W}}
\newcommand{\XXRM}{\mathrm{X}}
\newcommand{\YYRM}{\mathrm{Y}}

\newcommand{\ddRM}{\mathrm{d}}
\newcommand{\eeRM}{\mathrm{e}}

\newcommand{\rrRM}{\mathrm{r}}

\newcommand{\uuRM}{\mathrm{u}}

\newcommand{\yyRM}{\mathrm{y}}

\definecolor{Titleblue}{RGB}{0, 0, 100}

\newtheorem{assmpt}{Assumption}

\allowdisplaybreaks

\begin{document}

\begin{frontmatter}

\title{{Robust Stability of Discrete-time Disturbance Observers: Understanding Interplay of Sampling, Model Uncertainty and Discrete-time Designs}\thanksref{footnoteinfo}}

\thanks[footnoteinfo]{
This work was supported by the National Research Foundation of Korea (NRF) grant funded by the Korea government (Ministry of Science and ICT) (No. NRF-2017R1E1A1A03070342).	
The material of this paper was partially presented at the 54th IEEE Conference on Decision and Control, Osaka, Japan, December 15-18, 2015 \citep{Park2015}, and partially at the 15th International Conference on Control, Automation and Systems, Busan, Korea, October 13-16, 2015 \citep{Park2015a}.  
Corresponding author: H. Shim, Tel. +82-2-880-1745.}

\author[SNU]{Gyunghoon Park}\ead{gyunghoon.p@gmail.com},    % Add the 
\author[Hyundai]{Chanhwa Lee}\ead{chanhwa.lee@gmail.com},  % (ead) as shown
\author[UCF]{Youngjun Joo}\ead{Youngjun.Joo@ucf.edu},               % e-mail address 
\author[SNU]{Hyungbo Shim}\ead{hshim@snu.ac.kr} 

\address[SNU]{ASRI, Department of Electrical and Computer Engineering,, Seoul National University, Korea}  % Please supply                                              
\address[Hyundai]{Research \& Development Division, Hyundai Motor Company, Korea}  % Please supply                                              
\address[UCF]{Department of Electrical Engineering and Computer Science, University of Central Florida, USA}             % full addresses

                             % keyword list or with the 
                                          % help of the Automatica 
                                          % keyword wizard

\begin{abstract}                          % Abstract of not more than 200 words.
In this paper, we address the problem of robust stability for uncertain sampled-data systems controlled by a discrete-time disturbance observer (DT-DOB).
Unlike most of previous works that rely on the small-gain theorem, our approach is to investigate the location of the roots of the characteristic polynomial when the sampling is performed sufficiently fast.
This approach provides a generalized framework for the stability analysis in the sense that (i) many popular discretization methods are taken into account; (ii) under fast sampling, the obtained robust stability condition is necessary and sufficient except in a degenerative case; and (iii) systems of arbitrary order and of large uncertainty can be dealt with. 
The relation between sampling zeros---discrete-time zeros that newly appear due to the sampling---and robust stability is highlighted, and it is explicitly revealed that the sampling zeros can hamper stability of the overall system when the Q-filter and/or the nominal model are carelessly selected in discrete time.
Finally, a design guideline for the Q-filter and the nominal model in the discrete-time domain is proposed for robust stabilization under the sampling against the arbitrarily large (but bounded) parametric uncertainty of the plant.
\end{abstract}

\end{frontmatter}

\section{Introduction}

As a simple and effective tool for robust control, the disturbance observer (DOB) has received considerable attention over the past decades. 
Since its invention by \citet{Ohnishi87}, the DOB scheme has been successfully employed in many industrial problems, and in turn a large number of studies have been carried out to understand the underlying rationale behind its ability. 
(See \citet{SO2015,SPJBJ16,LYCC14} and the references therein.)

It is worth noting that a majority of the theoretical results on design and analysis of the DOB were developed in the continuous-time (CT) domain, while in practice the plant is usually controlled in the sampled-data fashion with a holder and a sampler. 
Thus even if a DOB is well-designed in the CT domain, one would face an additional problem of implementing the DOB in the discrete-time (DT) domain.
At first glance, this implementation issue may seem less important by the presumption that, as the sampling is performed sufficiently fast, any discretization of a CT-DOB will approximate its CT counterpart sufficiently well.
Yet interestingly, this is often not the case, and as reported in \citet{GHO02,BBS04,Uzunovic2018}, a naive conversion of the CT-DOB into the DT domain may significantly degrade the tracking performance or fail to stabilize the overall system in spite of sufficiently fast sampling.

\begin{figure}
	\begin{center}
		\includegraphics[width=.46\textwidth]{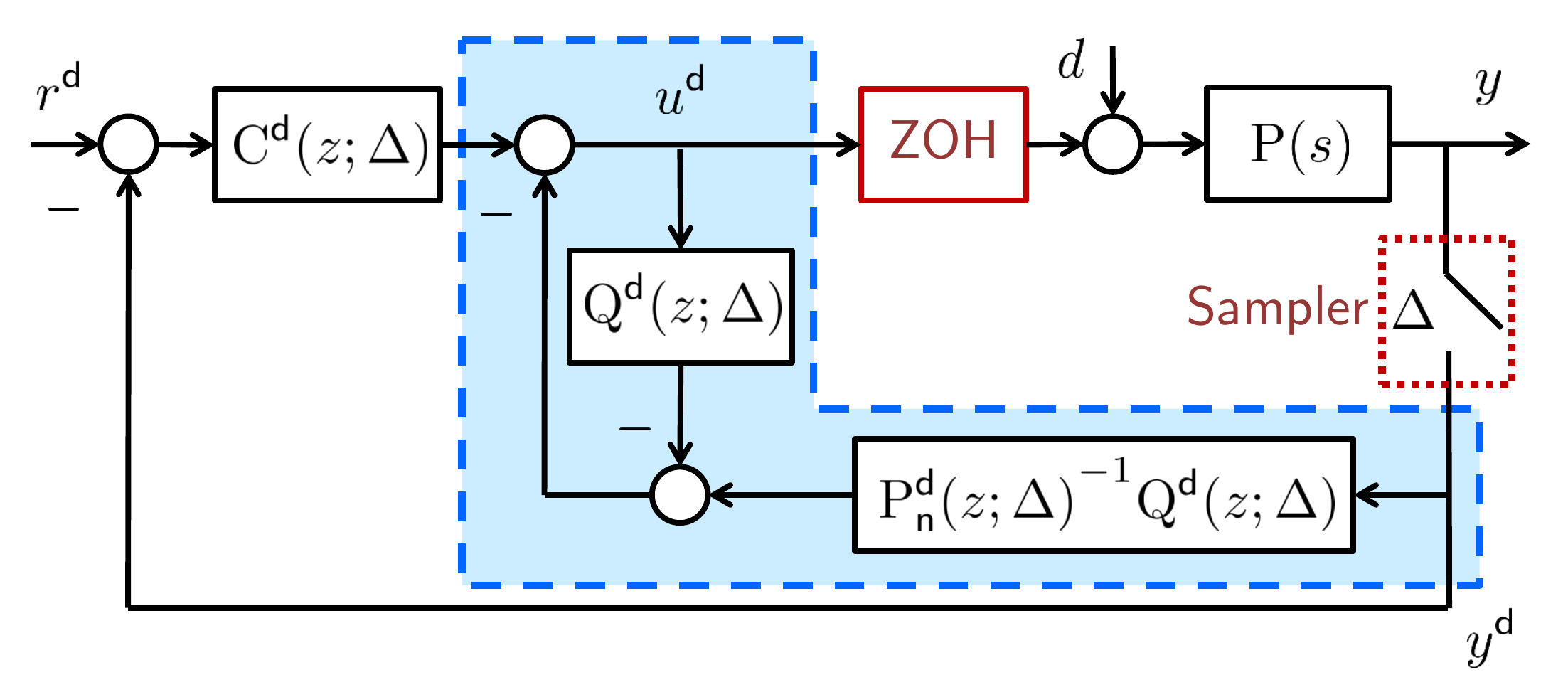}		
	\end{center}
	\caption{Overall configuration of sampled-data system controlled by DT-DOB (dotted block)}\label{fig:DT-DOB-DT-DOB}
\end{figure}

In this context, subsequent research efforts have been made to design a DT-DOB with a particular attention to the effect of the sampling on the overall stability.
While most relevant works share the common structure of the DT-DOB depicted in Fig.~\ref{fig:DT-DOB-DT-DOB} (or employ its equivalent blocks), there have been diverse ways of constructing the components of the DT-DOB; the DT nominal model {$\PPRM_{\mathsf n}^\ddSF$} and the DT Q-filter {$\QQRM^\ddSF$}.
On the one hand, the DT nominal model has been obtained by discretizing a CT nominal model of the actual plant in a number of methods.
A frequently used method for its discretization is the zero-order hold (ZOH) equivalence (or, exact discretization), because the ZOH is usually utilized in the sampled-data setting and thus any discretization methods other than the ZOH equivalence inevitably introduce additional model uncertainty to be compensated by the DOB \citep{TLT00,KK99}.
Yet this exact discretization often forces the inverse of the DT nominal model in the DOB loop to be unstable, because a ZOH equivalent model must have at least one unstable DT zero whenever the relative degree of the CT plant is larger than two and the sampling period is small.
To avoid this problem, approximate discretization methods have been preferred in some cases at the cost of discrepancy between the actual and nominal models in high-frequency range.
Possible candidates include the forward difference method \citep{LJS12}, the bilinear transformation \citep{YCC03,Lee1996}, the optimization-based techniques \citep{Chen2010,Kong2013}, and the adaptive algorithm-based approach \citep{CCKH16}. 
On the other hand, the design methodologies for the DT Q-filters presented in the literature can be categorized mainly into two groups.
One approach is to select a CT Q-filter a priori based on the CT-DOB theory, and to discretize it by taking into account the sampling process \citep{YCC03,Chen2010,TLT00,KK99,Lee1996}. 
Motivated by the works of \citet{GHO02,BBS04}, the discretization in this case aims to limit the bandwidth of the resulting DT Q-filter far below the Nyquist frequency, by which (roughly speaking) the DT-DOB does not conflict with the sampling process of the plant in a sense. 
Another stream of research dealt with the direct design of the Q-filter in DT domain \citep{KC03b,Chen2015,Kong2013,CCKH16}. 
In these works, a way of discretizing the CT nominal model was first determined, and then the DT Q-filter was constructed using the classical robust control theory developed for DT systems.

Despite increased interest, however, there is still a lack of understanding of the robust stability of the DT-DOB controlled systems in the sampled-data framework because of the following reasons.
First, the previous works listed above mainly employ the small-gain theorem or an approximation of the plant to be controlled in their stability analysis.
Thus (possibly conservative) sufficient conditions for robust stability were obtained, while it remains ambiguous how and when the sampling process leads to the instability of the DT-DOB controlled systems.
Next, the stability analysis was mostly performed ``after'' the discretization method for the DT nominal model is given, so that the effect of the selection of the DT nominal model on the overall stability is also unveiled.
Finally, many earlier works focused only on the second-order systems (e.g., mechanical systems) or on limited size of uncertainty (because of the use of small-gain theorem), which restricts the class of CT plants of interest. 

In this paper, we address the problem of robust stability for uncertain sampled-data systems controlled by DT-DOBs from a different perspective.
Unlike the previous works that rely on the small-gain theorem or the approximate discretization of the plant, the approach of this work is to express the DT-DOB in a general form and to investigate where the roots of the characteristic polynomial are located when the sampling proceeds sufficiently fast.
As a consequence, the proposed approach provides a generalized framework for the stability analysis in the sense that (i) various types of DT-DOB design methods in the relevant works are covered; (ii) under fast sampling, the stability condition is necessary and sufficient except in a degenerative case; and (iii) systems of arbitrary order and of large uncertainty can be dealt with.
This is done in Sections~\ref{sec:DT-DOB-DT-DOB} and \ref{sec:DT-DOB-Stability} on the basis of the review on the sampled-data systems in Section~\ref{sec:SDS-Basic}.

This work confirms that there exists a strong relation between the sampling process, the design of the DT-DOB, and the uncertainty of the plant in the stability issue.
To gain further insight into the role of the sampling process, we re-interpret the rules of thumb that have been referred to in the usual DT-DOB designs mentioned above.
Our results highlight that the sampling zeros (i.e., extra DT zeros of the sampled-data model that are generated by the sampling process) may hamper the stability of the overall system unless the DT Q-filter and the discretization method for the DT nominal model are carefully selected (Section~\ref{sec:Remark}).
To tackle the stability issues caused by the sampling process as well as to ensure robust stability against arbitrarily large but bounded parametric uncertainty, in Section~\ref{sec:DT-DOB-Design} we provide a new systematic guideline for the direct design of the DT-DOB based on the theoretical result.
In Section~\ref{sec:DT-DOB-Sim}, it is shown through some simulations that, compared with a discretization of a CT-DOB, the proposed design guideline in this paper can allow relatively larger bandwidths of the DT Q-filter with stability guaranteed, which ensures better disturbance rejection performance.

{\bf Notation}: We denote the coefficient of $s^m$ in the polynomial $(s+1)^n$ as $\left( \begin{smallmatrix}
n\\
m
\end{smallmatrix} \right)$.
For a CT signal $a(t)$, $\mathcal{L}\{ a(t)\}$ stands for the Laplace transform of $a(t)$. 
Similarly, the $\mathcal{Z}$-transform of a DT signal $a^\ddSF[k]$ is denoted by $\mathcal{Z}\{a^\ddSF[k]\}$.
(Hereinafter, we shall use the superscript ``$\ddSF$'' to indicate that a parameter or a variable is associated with the DT domain.)
The sets of the real numbers, the positive real numbers, and the complex numbers are denoted by $\mathbb{R}$, $\mathbb{R}_{>0}$, and $\mathbb{C}$, respectively. 
For two vectors $a$ and $b$, we use $[a;b]$ to represent $[a^\top,b^\top]^\top$.
The symbols ${\rm Re}(\xi^\star)$ and ${\rm Im}(\xi^\star)$ indicate the real and the imaginary parts of $\xi^\star \in \mathbb{C}$, respectively. 
If all the coefficients of a polynomial $\NNRM(z)$ (and a rational function $\PPRM(z)$) are functions of a variable $\Delta$, the polynomial is usually written as $\NNRM(z;\Delta)$ instead of $\NNRM(z)$ (and similarly, the rational function as $\PPRM(z;\Delta)$ rather than $\PPRM(z)$).
For two sets $\mathcal{V}$ and $\mathcal{W}$,  ${\mathbf{C}}_{\mathcal{W}}({\mathcal{V}})$ stands for the set of continuous functions from $\mathcal{V}$ to $\mathcal{W}$. 
To shorten the notation, let ${\mathbf C}_{\mathbb R}:= {\mathbf C}_{\mathbb R}(\mathbb{R}_{>0})$.
The set of polynomials of $z$ with coefficients in $\mathbb{R}$ is denoted by $\mathbb{R}[z]$.
In a similar manner, we use the notation ${\mathbf C}_{\mathbb R}[z]$ to indicate the set of polynomials (e.g., $\NNRM(z;\Delta) = \NNRM_n(\Delta)z^n + \cdots + \NNRM_1(\Delta)z + \NNRM_0(\Delta)$) whose coefficients are in $\mathbf C_{\mathbb R}$.

\section{Problem Formulation}\label{sec:DT-DOB-SDS}

Consider a single-input single-output (SISO) continuous-time (CT) plant 
\begin{align}\label{eq:DT-DOB-yFreqCT}
\yyRM(s) = \PPRM(s) \big(\uuRM(s) + \ddRM(s)\big)
\end{align}
where $\uuRM(s)$ is the input, $\yyRM(s)$ is the output, $\ddRM(s)$ is the disturbance, and 
\begin{equation}\label{eq:DT-DOB-Plant}
\PPRM (s)=\frac{g\prod_{i=1}^{n-\nu}\left(s-\zzSF_{i}\right)}{\prod_{i=1}^{n}(s-\ppSF_{i})}=:\frac{\NNRM (s)}{\DDRM (s)}
\end{equation}
where $\nu\geq 1$ is the relative degree of $\PPRM(s)$, and $\zzSF_{i}\in \mathbb{C}$, $\ppSF_{i} \in \mathbb{C}$, and $g \neq 0$ are the CT zeros, the CT poles, and the high-frequency gain of $\PPRM(s)$, respectively.
Without loss of generality, assume that the polynomials $\NNRM(s)$ and $\DDRM(s)$ are coprime.
It is also supposed that the plant \eqref{eq:DT-DOB-Plant} has bounded parametric uncertainty of arbitrarily large size and the sign of the high-frequency gain $g$ is known, as stated below.

\begin{assmpt}\label{asm:DT-DOB-Uncertainty}
	The CT plant $\PPRM (s)$ in \eqref{eq:DT-DOB-Plant} is contained in the set of uncertain transfer functions
	\begin{align}
	& \mathcal{P}:=\bigg\{ g\frac{s^{n-\nu} + \beta_{n-\nu-1} s^{n-\nu-1} +  \cdots + \beta_0}{s^n + \alpha_{n-1}s^{n-1}+\cdots + \alpha_0}:\label{eq:DT-DOB-SetP} \\
	&\qquad ~~~~ 0<\underline{g} \leq g \leq \overline{g}, ~~ \underline{\alpha}_i \leq \alpha_i \leq \overline{\alpha}_i, ~~ \underline{\beta}_i \leq  \beta_i \leq \overline{\beta}_i  \bigg\}\notag
	\end{align}
	where the bounds $\underline{g}$, $\overline{g}$, $\underline{\alpha}_i$, $\overline{\alpha}_i$, $\underline{\beta}_i$, and $\overline{\beta}_i$ are known. 
	$\hfill\square$
\end{assmpt}

In this paper, we are interested in the situation when the CT plant \eqref{eq:DT-DOB-yFreqCT} is controlled in the sampled-data framework, together with the following two components.
\begin{itemize}
	\item {\bf Zero-order hold} (ZOH): $u(t)=u^{\mathsf d}[k]$ for $k\Delta \leq t < (k+1)\Delta$ where $\Delta>0$ is the sampling period,  $u(t)$ is the CT input of the plant (i.e., $u(t)=\mathcal{L}^{-1}\{{\uuRM(s)}\}$), and $u^{\mathsf d}[k]$ is the DT control input generated by a DT controller. 
	\item {\bf Sampler}: $y^{\mathsf d}[k] = y(k\Delta)$ in which $y(t)$ is the CT output of the plant (i.e., $y(t)=\mathcal{L}^{-1}\{\yyRM(s)\}$), and $y^{\mathsf d}[k]$ indicates the DT output to be measured at each sampling time. 
\end{itemize}

As a particular DT control scheme, this paper considers the DT disturbance observer (DT-DOB).
Overall configuration of the DT-DOB-based controller (DT-DOBC) is depicted in Fig.~\ref{fig:DT-DOB-DT-DOB}.
In the figure, the DT transfer functions $\QQRM^\ddSF(z;\Delta)$, $\PPRM_\nnSF^\ddSF(z;\Delta)$, and $\CCRM^\ddSF(z;\Delta)$ (whose coefficients are possibly dependent of $\Delta$) stand for the DT Q-filter, the DT nominal model, and the DT nominal controller, respectively. 
The DT signal $r^\ddSF[k]$ is the reference command for the sampled output $y^\ddSF[k]$.

The main purpose of this paper is to derive a robust stability condition for the DT-DOB controlled sampled-data systems. 
In particular, we aim to clarify how the sampling process (as well as the DT-DOB design and the parametric uncertainty of the plant) influences the stability of the overall system, and to express their relation in an explicit manner.
Based on the stability analysis, it is also desired to provide a systematic design guideline for the DT-DOBC to ensure robust stabilization against both sampling process and parametric uncertainty.

\section{Basics on Sampled-data Systems}\label{sec:SDS-Basic}

This section summarizes some inherent natures of the sampled-data system. 
Let us begin with a minimal realization of the CT plant \eqref{eq:DT-DOB-yFreqCT} in the state space given by
\begin{subequations}\label{eq:DT-DOB-CTplantSS}
\begin{align}
\dot{x}(t) & = A x(t) + B \big(u(t)+d(t)\big), \\
y(t) & = C x(t)
\end{align} 
\end{subequations}
in which the matrices $A$, $B$, and $C$ satisfy $\PPRM(s) = C(sI-A)^{-1}B$.
Then the corresponding sampled-data system with the ZOH and the sampler can be written in the DT domain by
\begin{subequations}
	\begin{align}\label{eq:DT-DOB-PlantSSDT}
	x^\ddSF[k+1] & = A^\ddSF {x}^\ddSF[k] + B^\ddSF u^\ddSF[k] + \hat{d}^\ddSF[k],\\
	y^\ddSF[k] & = C^\ddSF x^\ddSF[k]
	\end{align}
\end{subequations}
where $x^\ddSF[k] := x(k\Delta)$ is the DT state, $A^\ddSF(\Delta):= e^{A\Delta}$, $B^\ddSF(\Delta):= \int^{\Delta}_0 e^{A \rho} B {\rm d}\rho$, $C^\ddSF := C$, and the DT disturbance $\hat{d}^\ddSF[k]$ is of the form
\begin{align}\label{eq:DT-DOB-DTdist}
\hat{d}^\ddSF[k]:= \int^{(k+1)\Delta}_{k\Delta} e^{A((k+1)\Delta - \rho)} B d(\rho) {\rm d}\rho \in {\mathbb{R}}^n.
\end{align}
For each $\PPRM\in \mathcal{P}$, let $\mathcal{T}_{{\mathrm P},1}$ be the set of $\Delta\in \mathbb{R}_{>0}$ such that $\Delta(\lambda - \lambda') = 2\pi j k$ holds for a nonzero integer $k$ and some distinct eigenvalues $\lambda$ and $\lambda'$ of $A$. 
One can readily see that each $\mathcal{T}_{{\PPRM},1}$ has the measure zero and $\inf\{ \mathcal{T}_{\PPRM,1} \}>0$.
It is well known in the literature that if the triplet $(A,B,C)$ is controllable and observable, then the corresponding  $(A^\ddSF(\Delta),B^\ddSF(\Delta), C^\ddSF )$ is also controllable and observable for all  $\Delta\in \mathbb{R}_{>0} \setminus \mathcal{T}_{\PPRM,1}$.

Now define ${\uuRM}^\ddSF (z):= \mathcal{Z}\{u^\ddSF [k]\}$, $\yyRM^\ddSF (z):= \mathcal{Z} \{ y^\ddSF[k]\}$, and $\hat{\ddRM}^\ddSF(z):=\mathcal{Z}\{ \hat{d}^\ddSF[k]\}$.
Then we obtain a frequency-domain expression of the sampled-data system
\begin{align}
{\yyRM}^\ddSF(z) = \PPRM^\ddSF (z;\Delta) {\uuRM}^\ddSF (z)+ {\WWRM}^\ddSF (z;\Delta) \hat{\ddRM}^\ddSF (z)\label{eq:DT-DOB-yz}
\end{align}
where $\PPRM^\ddSF(z;\Delta) := C^\ddSF (zI-A^\ddSF(\Delta))^{-1} B^\ddSF(\Delta) = \mathcal{Z}\{ \mathcal{L}^{-1}\{(1-e^{-\Delta s})/{s}\times \PPRM (s)\}|_{t=k\Delta}\}$ and ${{\WWRM}}^\ddSF(z;\Delta) := C^\ddSF (zI-A^\ddSF(\Delta))^{-1}$. 
Since $\PPRM^\ddSF(z;\Delta)$ and $\WWRM^\ddSF(z;\Delta)$ have the same denominator $\DDRM^\ddSF(z;\Delta):=\det(zI-A^\ddSF(\Delta))$, they
can be expressed as $\PPRM^\ddSF(z;\Delta) = \NNRM^\ddSF(z;\Delta)/\DDRM^\ddSF(z;\Delta)$ and $\WWRM^\ddSF(z;\Delta) = (1/\DDRM^\ddSF(z;\Delta)) [\NNRM^\ddSF_{\wwSF,1}(z;\Delta),\dots, \NNRM^\ddSF_{\wwSF,n}(z;\Delta)]$ with some polynomials $\NNRM^\ddSF\in {\mathbf C}_{\mathbb{R}}[z]$ and $\NNRM^\ddSF_{{\mathsf w},i}\in {\mathbf C}_{\mathbb{R}}[z]$, $i=1,\dots,n$.
In addition, for all $\Delta\in \mathbb{R}_{>0} \setminus \mathcal{T}_{\PPRM,1}$, there is no pole-zero cancellation in both $\PPRM^\ddSF(z;\Delta)$ and $\WWRM^\ddSF(z;\Delta)$ (so that the roots of $\DDRM^\ddSF (z;\Delta)$ do not coincide with any roots of $\NNRM^\ddSF(z;\Delta)$ nor $\NNRM^\ddSF_{{\mathsf w},i}(z;\Delta)$).

In the remainder of this section, we discuss the DT transfer function $\PPRM^\ddSF(z;\Delta)$ in \eqref{eq:DT-DOB-yz} from $\uuRM^\ddSF(z)$ to $\yyRM^\ddSF(z)$. 
Since $\PPRM^\ddSF(z;\Delta)$ ``exactly'' represents the input-to-output relation of the sampled-data system with the ZOH, it is often called the {\emph{ZOH equivalent model}} of $\PPRM (s)$ \citep{AHS84}. 
The following lemma describes a well-known fact on the ZOH equivalent model.
\begin{lem}\label{lem:DT-DOB-ZOH} \citep[Theorem~1]{AHS84}, \citep[Lemma~5.10]{YG14}
	For each $\PPRM\in \mathcal{P}$, let ${\mathcal{T}}_{\PPRM,2}:= \{ \Delta\in \mathbb{R}_{>0}: C^\ddSF B^\ddSF(\Delta) = 0 \}$. 
	Then $\mathcal{T}_{\PPRM,2}$ is a measure zero set and the ZOH equivalent model $\PPRM^\ddSF(z;\Delta)$ of $\PPRM (s)$ has the relative degree 1 for all $\Delta\in {\mathbb R}_{>0} \setminus \mathcal{T}_{\PPRM,2}$.
	More precisely, for all $\Delta\in {\mathbb R}_{>0} \setminus \mathcal{T}_{\PPRM,2}$, $\PPRM^\ddSF(z;\Delta)$ has the form
	\begin{align}
	\PPRM^\ddSF(z;\Delta) & =  \frac{\NNRM^\ddSF(z;\Delta)}{\DDRM^\ddSF (z;\Delta)}  = 
	\frac{g  \MMRM^\ddSF (z;\Delta) { \prod_{i=1}^{n-\nu} \big({z- \zzSF_i^\ddSF(\Delta)}\big)/{\Delta}}}{  \prod_{i=1}^{n} \big({z- \ppSF_i^\ddSF(\Delta)}\big)/{\Delta}} \label{eq:DT-DOB-ZOHEx}
	\end{align}
	where $\zzSF_i^\ddSF, \ppSF_i^\ddSF \in {\mathbf{C}}_{\mathbb{C}}\big(\mathbb{R}_{>0}\setminus\mathcal{T}_{\PPRM,2}\big)$ and $\MMRM^\ddSF \in {\mathbf C}_{\mathbb{R}}[z]$ is a polynomial of order $\nu-1$. 
	In addition, as $\Delta\rightarrow 0^+$, 
	\begin{itemize}
		\item $\ppSF_{i}^\ddSF (\Delta)\rightarrow 1$ for $i=1,\cdots,n$,
		\item $\zzSF_{i}^\ddSF (\Delta)\rightarrow 1$ for $i=1,\cdots,n-\nu$,
		\item $\MMRM^\ddSF (z;\Delta)\rightarrow {\BBRM}_{\nu-1}(z)/\nu!=: {\MMRM}^\star(z)$
	\end{itemize}
	where ${\BBRM}_{\nu-1}(z)  := b_{(\nu-1, \nu-1)} z^{\nu-1}+\cdots+b_{(\nu-1, 0)}$ is the Euler-Frobenius polynomial of order $\nu-1$, with the coefficients
	${b}_{(\nu-1,j)}:= \sum^{\nu-j}_{l=1}(-1)^{\nu-j-l} l^\nu
	\left(\begin{smallmatrix}
	\nu+1\\
	\nu-j-l
	\end{smallmatrix}\right)$ for $j=0,\dots,\nu-1$. 	
\end{lem}

Among the DT zeros in \eqref{eq:DT-DOB-ZOHEx}, $\zzSF_i^\ddSF(\Delta)$ are associated with the CT zeros of $\PPRM(s)$ in the sense that each can be approximated by $e^{\zzSF_i \Delta}$ with sufficiently small $\Delta$ \citep{AHS84,YG14}. 
For this reason, we call $\zzSF_i^\ddSF(\Delta)$ as the {\it intrinsic zeros} of $\PPRM^\ddSF(z;\Delta)$.
On the other hand, the other zeros (i.e., the roots of ${\MMRM^\ddSF (z;\Delta)}=0$) newly appear due to the sampling process, and so they are often named the {\it sampling zeros}.
The above lemma also points out that, unlike the intrinsic zeros, the limit of the sampling zeros is solely determined by the Euler-Frobenius polynomial ${\BBRM}_{\nu-1}(z)$, which is independent of the plant's characteristics. 
Some important properties of the polynomial are summarized as follows.

\begin{lem}\label{lem:DT-DOB-EF} \citep{AHS84,YG14}	
	The Euler-Frobenius polynomial ${\BBRM}_{\nu-1}(z)$ in Lemma~\ref{lem:DT-DOB-ZOH} satisfies the following statements:
	\begin{enumerate}[\hspace{3ex}]		
		\item[(a)] $b_{(\nu-1,i)} = b_{(\nu-1, \nu-1-i)}$ for all $i=0,\dots,\nu-1$, and  $b_{(\nu-1,0)} = b_{( \nu-1, \nu-1) }=1$;
		\item[(b)] ${\BBRM}_{\nu-1}(1) = \nu!$;
		\item[(c)] For $\nu\geq 3$, there is at least one root of ${\BBRM}_{\nu-1}(z)=0$ outside the unit circle;
		\item[(d)] All the roots of ${\BBRM}_{\nu-1}(z)=0$ are single and negative real.
	\end{enumerate}
\end{lem}

A natural consequence from Item (a) is that $\BBRM_{\nu-1}(\zeta)=0$  for some $\zeta\in \CCBB$ implies $\BBRM_{\nu-1}(\zeta^{-1})=0$, and that $\BBRM_{\nu-1}(-1)=0$ when $\nu$ is even.
It is obtained from Lemmas~\ref{lem:DT-DOB-ZOH} and~\ref{lem:DT-DOB-EF} that with high relative degree $\nu$ of $\PPRM (s)$ (that is, $\nu\geq 3$) and fast sampling, the sampled-data model $\PPRM^\ddSF(z;\Delta)$ is ``inherently'' of non-minimum phase in the DT domain (even in the case when the corresponding CT plant is of minimum phase). 
We will show shortly that this nature of the sampled-data model incurs a significant difference between the stability conditions for the CT- and the DT-DOB schemes in the end. 

For further analysis, the following lemma on the uncertain sampled-data systems is needed.
\begin{lem}\label{lem:DeltaStar}
	${\Delta}_{\mathcal{P}}^\star:={\rm inf}\big\{\bigcup\nolimits_{\PPRM \in {\mathcal P}}{{\mathcal T}}_{\PPRM} \big\}$ is nonzero where ${\mathcal T}_{\PPRM}:= {\mathcal T}_{\PPRM,1} \bigcup {\mathcal T}_{\PPRM,2}$ and $\mathcal{P}$ is given in Assumption~\ref{asm:DT-DOB-Uncertainty}.
\end{lem}

{\bf PROOF.} The proof is provided in Appendix. 
$\hfill\blacksquare$

According to the above lemma, it can be concluded that the explicit form \eqref{eq:DT-DOB-ZOHEx} of $\PPRM^\ddSF(z;\Delta)$ is valid for all $\Delta\in (0,\Delta^\star_{\mathcal P})$ and for all the CT plants \eqref{eq:DT-DOB-Plant} in $\mathcal{P}$, without any (unstable) pole-zero cancellation.

\section{General Expression of Discrete-time Disturbance Observer-based Controllers}\label{sec:DT-DOB-DT-DOB}

As an intermediate step, this section introduces explicit forms of $\QQRM^\ddSF(z;\Delta)$, $\PPRM_\nnSF^\ddSF(z;\Delta)$, and $\CCRM^\ddSF(z;\Delta)$ comprising the DT-DOBC in Fig.~\ref{fig:DT-DOB-DT-DOB}.
It should be noted first that most of the previous analyses are based on a specific design of the DT-DOB, which possibly brings conservatism.  
In this context, a particular interest here is to present a ``general expression'' of the DT-DOBC, in the sense that a variety of design methodologies for the DT-DOB can be dealt with at once.

We start with the DT nominal model $\PPRM_\nnSF^\ddSF(z;\Delta)$ and the DT nominal controller $\CCRM^\ddSF(z;\Delta)$.
The usual way to obtain them is to discretize a CT nominal model
\begin{subequations}
	\begin{equation}\label{eq:DT-DOB-CTNominalModel}
	\PPRM_{\nnSF} (s) = \frac{g_{\nnSF}\prod_{i=1}^{n-\nu}(s-\zzSF_{{\nnSF},i})}{\prod_{i=1}^{n}(s-\ppSF_{{\nnSF},i})}=:\frac{\NNRM_{\nnSF}(s)}{\DDRM_{\nnSF}(s)}
	\end{equation}
	of $\PPRM (s)$ and a CT nominal controller 
	\begin{equation}
	\CCRM(s) = \frac{g_{\ccSF}\prod^{n_{\ccSF}-\nu_{\ccSF}}_{i=1}(s-\zzSF_{{\ccSF},i})}{\prod^{n_{\ccSF}}_{i=1}(s-\ppSF_{{\ccSF},i})}=:\frac{\NNRM_{\ccSF}(s)}{\DDRM_{\ccSF}(s)},\label{eq:DT-DOB-CTController}
	\end{equation}
\end{subequations}
in which $g_\nnSF > 0$ and $g_\ccSF \neq 0$ are the high-frequency gains, $\zzSF_{\nnSF,i}$ and $\zzSF_{\ccSF,i}$ are the CT zeros, $\ppSF_{\nnSF,i}$ and $\ppSF_{\ccSF,i}$ are the CT poles of $\PPRM_{\nnSF} (s)$ and $\CCRM(s)$, respectively.
It is assumed that $\PPRM_{\nnSF}(s)$ belongs to the set $\mathcal{P}$ in \eqref{eq:DT-DOB-SetP}, and $\CCRM(s)$ is selected such that there is no unstable pole-zero cancellation in $\PPRM_{\nnSF}(s) \CCRM(s)/(1+\PPRM_{\nnSF}(s) \CCRM(s))$. 
At this stage, we should emphasize that even if $\PPRM_\nnSF(s)$ and $\CCRM(s)$ are fixed, their discretized transfer functions $\PPRM^\ddSF_\nnSF(z;\Delta)$ and $\CCRM^\ddSF(z;\Delta)$ may not be unique, as several discretization methods are available.
Keeping this in mind, the following assumption is made in order to represent a large number of the discretization results into a unified form.
(To proceed, we represent $\PPRM^\ddSF_\nnSF(z;\Delta)$ and $\CCRM^\ddSF(z;\Delta)$ as the ratios of two polynomials: $\PPRM^\ddSF_\nnSF(z;\Delta) = \NNRM^\ddSF_\nnSF(z;\Delta)/\DDRM^\ddSF_\nnSF(z;\Delta)$ and $\CCRM^\ddSF(z;\Delta) = \NNRM^\ddSF_\ccSF(z;\Delta)/\DDRM^\ddSF_\ccSF(z;\Delta)$.)

\begin{assmpt}\label{asm:DT-DOB-DTC}
	The DT transfer functions $\PPRM^\ddSF_\nnSF(z;\Delta)$ and $\CCRM^\ddSF(z;\Delta)$ are such that  $\NNRM^\ddSF_\nnSF, \NNRM^\ddSF_\ccSF,\DDRM^\ddSF_\nnSF,\DDRM^\ddSF_\ccSF \in {\mathbf C}_{\mathbb{R}}[z]$. 
	Moreover, there is a measure zero set  $\mathcal{T}_{\nnSF\ccSF}\subset \mathbb{R}_{>0}$ such that  $\Delta^\star_{\nnSF\ccSF}:={\rm inf}\{\TTCAL_{\nnSF\ccSF}\}>0$, and for all $\Delta \in \mathbb{R}_{> 0 }\setminus\mathcal{T}_{\nnSF\ccSF}$, $\PPRM^\ddSF_\nnSF(z;\Delta)$ and $\CCRM^\ddSF(z;\Delta)$ have the form
	\begin{subequations} 
		\begin{align}
		\PPRM_{\nnSF}^\ddSF (z;\Delta) & =  \frac{\NNRM_{\nnSF}^\ddSF(z;\Delta)}{\DDRM_{\nnSF}^\ddSF(z;\Delta)}\label{eq:DT-DOB-DTNominalModel}\\
		& = \frac{g_{\nnSF}^\ddSF(\Delta) \MMRM_{\nnSF}^\ddSF(z;\Delta) \prod_{i=1}^{n-\nu} \big({z-\zzSF_{{\nnSF},i}^\ddSF(\Delta)}\big)/{\Delta}}{ \prod_{i=1}^{n} \big({z-\ppSF_{{\nnSF},i}^\ddSF(\Delta)}\big)/{\Delta}},\notag\\
		\CCRM^\ddSF(z;\Delta) & =  \frac{\NNRM_{\ccSF}^\ddSF(z;\Delta)}{\DDRM_{\ccSF}^\ddSF(z;\Delta)} \label{eq:DT-DOB-DTController} \\
		& =  \frac{g_\ccSF^\ddSF(\Delta) \MMRM_{\ccSF}^\ddSF (z;\Delta) \prod^{n_{\ccSF}-\nu_{\ccSF}}_{i=1} \big({z-\zzSF_{{\ccSF},i}^\ddSF(\Delta)}\big)/{\Delta}}{ \prod^{n_{\ccSF}}_{i=1}
			\big({z-\ppSF_{{\ccSF},i}^\ddSF(\Delta)}\big)/{\Delta}}\notag 
		\end{align}
	\end{subequations}	 
	where $\MMRM^\ddSF_{\nnSF}, \MMRM^\ddSF_{\ccSF}\in {\mathbf C}_{\mathbb{R}}[z]$ are polynomials of order $n_{\mmSF\nnSF}\leq {\nu}$ and $ n_{\mmSF\ccSF} \leq \nu_\ccSF$, respectively,   
	$g_\nnSF^\ddSF, g_\ccSF^\ddSF \in {\mathbf{C}}_{\mathbb{R}}(\mathbb{R}_{>0}\setminus \mathcal{T}_{\nnSF\ccSF})$, and $\zzSF_{\nnSF,i}^\ddSF, \zzSF_{\ccSF,i}^\ddSF, \ppSF_{\nnSF,i}^\ddSF, \ppSF_{\ccSF,i}^\ddSF \in {\mathbf{C}}_{\mathbb{C}}(\mathbb{R}_{>0}\setminus \mathcal{T}_{\nnSF\ccSF})$.
	As $\Delta\rightarrow 0^+$,
	\begin{subequations}
		\begin{align}
		g^\ddSF_\nnSF(\Delta) & \rightarrow g_\nnSF, & g^\ddSF_\ccSF(\Delta)&\rightarrow g_\ccSF,\\
		\MMRM^\ddSF_{\nnSF}(z;\Delta) & \rightarrow \MMRM_{\nnSF}^{\star}(z), & \MMRM^\ddSF_{\ccSF}(z;\Delta) & \rightarrow  \MMRM_{\ccSF}^{\star}(z),\\
		\frac{\zzSF_{{\nnSF},i}^\ddSF(\Delta)-1}{\Delta} & \rightarrow \zzSF_{{\nnSF},i}, & \frac{\zzSF_{{\ccSF},i}^\ddSF(\Delta)-1}{\Delta} & \rightarrow \zzSF_{{\ccSF},i},\\
		\frac{\ppSF_{{\nnSF},i}^\ddSF(\Delta)-1}{\Delta} & \rightarrow \ppSF_{{\nnSF},i}, & \frac{\ppSF_{{\ccSF},i}^\ddSF(\Delta)-1}{\Delta} & \rightarrow \ppSF_{{\ccSF},i}
		\end{align}
	\end{subequations}
	where $\MMRM_{\nnSF}^{\star}, \MMRM_{\ccSF}^\star \in {\mathbf C}_{\mathbb{R}}[z]$ are polynomials of order $n_{\mmSF\nnSF}$ and $n_{\mmSF\ccSF}$ satisfying $\MMRM_{\nnSF}^{\star}(1)=1$ and $\MMRM_{\ccSF}^{\star}(1)=1$, respectively.	
\end{assmpt}

\begin{table*}[t]
	\centering
	{
		\begin{tabular}{>{\centering\arraybackslash} m{1.7cm}
				>{\centering\arraybackslash}m{1.35cm}
				>{\centering\arraybackslash}m{2.8cm}
				>{\centering\arraybackslash}m{3.2cm}
				>{\centering\arraybackslash}m{3.2cm}
			}
			\toprule
			& {FDM} & BDM & BT & MPZ  \\
			\midrule
			{$g_{\nnSF}^\ddSF(\Delta)$} & $g_{\nnSF}$ & $\begin{aligned}g_{\nnSF}\frac{\prod_{i=1}^{n-\nu}(1-\Delta \zzSF_{{\nnSF},i})}{\prod_{i=1}^{n}(1-\Delta \ppSF_{{\nnSF},i})}\end{aligned}$ & $\begin{aligned}g_{\nnSF}\frac{\prod_{i=1}^{n-\nu}\bigg(1-\dfrac{\Delta \zzSF_{{\nnSF},i}}{2}\bigg)}{\prod_{i=1}^{n}\bigg(1-\dfrac{\Delta \ppSF_{{\nnSF},i}}{2}\bigg)}\end{aligned}$ & $\begin{aligned}g_{\nnSF}\frac{\prod_{i=1}^{n}\dfrac{e^{\Delta\ppSF_{{\nnSF},i}}-1}{\Delta\ppSF_{{\nnSF},i}}}{\prod_{i=1}^{n-\nu}\dfrac{e^{\Delta\zzSF_{{\nnSF},i}}-1}{\Delta\zzSF_{{\nnSF},i}}}\end{aligned}$\\
			\midrule        
			{$\MMRM_{\nnSF}^\ddSF(z;\Delta)$}  & $1$ & $z^{\nu}$ & $\dfrac{(z+1)^\nu}{2^\nu} $ & $\begin{aligned}\frac{(z+1)^\nu}{2^\nu}\end{aligned}$ \\ 
			\midrule
			{$\zzSF^\ddSF_{{\nnSF},i}(\Delta)$}& $1+\Delta \zzSF_{{\nnSF},i}$ & $\begin{aligned}\frac{1}{1- \Delta \zzSF_{{\nnSF},i}}\end{aligned}$ & $\begin{aligned}\frac{1+ \dfrac{\Delta\zzSF_{{\nnSF},i}}{2}}{1- \dfrac{\Delta  \zzSF_{{\nnSF},i}}{2}}\end{aligned}$ & $e^{\Delta \zzSF_{{\nnSF},i}}$\\
			\midrule
			{$\ppSF^\ddSF_{{\nnSF},i}(\Delta)$}& $1+\Delta \ppSF_{{\nnSF},i}$ & ${\begin{aligned}\frac{1}{1-\Delta \ppSF_{{\nnSF},i}}\end{aligned}}$ & $\begin{aligned}\frac{1+\dfrac{\Delta \ppSF_{{\nnSF},i}}{2}}{1-\dfrac{\Delta \ppSF_{{\nnSF},i}}{2}}\end{aligned}$ & $e^{\Delta \ppSF_{{\nnSF},i}}$\\
			\bottomrule
	\end{tabular}}
	\vspace{0.15cm}
	\caption{Components of DT nominal models $\PPRM_\nnSF^\ddSF(z;\Delta)$ in \eqref{eq:DT-DOB-DTNominalModel} obtained from typical discretization methods; forward difference method (FDM), backward difference method (BDM), bilinear transformation (or Tustin's method, BT), and matched pole-zero method (MPZ).}
	\label{tab:Discretizations}
\end{table*}

Roughly speaking, the statement of Assumption~\ref{asm:DT-DOB-DTC} is an extension of that in Lemma~\ref{lem:DT-DOB-ZOH}, from the ZOH equivalence method to other approximate discretization methods widely employed in the literature. 
As a matter of fact, Assumption~\ref{asm:DT-DOB-DTC} is satisfied with the forward difference method (FDM, $s=(z-1)/\Delta$), the backward difference method (BDM, $s=(z-1)/(\Delta z)$), the bilinear transformation (BT, or Tustin's method, $s=(2(z-1))/(\Delta(z+1))$), and the matched pole zero method (MPZ) \citep{FPW98}, as summarized in Table~\ref{tab:Discretizations}. 
(We note that the above approximate methods result in $\TTCAL_{\nnSF\ccSF} = \emptyset$.) 

On the other hand, as usual in the literature, the DT Q-filter $\QQRM^\ddSF(z;\Delta)$ in Fig.~\ref{fig:DT-DOB-DT-DOB} is selected as a stable DT low-pass filter.
Particularly, in order to gain additional design freedom, it is further supposed that the coefficients of $\QQRM^\ddSF(z;\Delta)$ are possibly dependent of the sampling period $\Delta$. 
Thus $\QQRM^\ddSF(z;\Delta)$ is of the form
\begin{align}
&\QQRM^\ddSF(z;\Delta)  = \frac{\NNRM_{\qqSF}^\ddSF(z;\Delta)}{\DDRM_{\qqSF}^\ddSF(z;\Delta)} \label{eq:DT-DOB-Q} \\
 & \quad~~ := \frac{c_{m_\qqSF}^\ddSF (\Delta)(z-1)^{m_\qqSF}+\cdots+c_0^\ddSF(\Delta)}{(z-1)^{n_\qqSF}+{a}_{n_\qqSF-1}^\ddSF(\Delta)(z-1)^{n_\qqSF-1}+\cdots+{a}_0^\ddSF(\Delta)}\notag
\end{align}
where $c_i^\ddSF, a^\ddSF_i \in \mathbf{C}_{\mathbb{R}}(\mathbb{R}_{>0})$ satisfying that $c_i^\ddSF(\Delta)\rightarrow c_i^\star$ and ${a}_i^\ddSF(\Delta)\rightarrow {a}_i^\star$ as $\Delta\rightarrow 0^+$, $c_0^\ddSF(\Delta) = {a}_0^\ddSF(\Delta)$ for all $\Delta\in \mathbb{R}_{> 0}$, and $c_0^\star = {a}_0^\star \neq 0$.
(The second condition is needed for the unity DC gain.)
The degrees $n_\qqSF$ and $m_\qqSF$ of the denominator and the numerator of $\QQRM^\ddSF(z;\Delta)$ are chosen to satisfy $n_\qqSF - m_\qqSF \geq \max\{\nu-n_{\mmSF\nnSF},1\}$, by which the block ${\PPRM^\ddSF_\nnSF}(z;\Delta)^{-1}\QQRM^\ddSF(z;\Delta)$ in Fig.~\ref{fig:DT-DOB-DT-DOB} is {proper (and thus implementable) and $\QQRM^\ddSF(z;\Delta)$ itself is strictly proper.

At last, we conclude this section by reminding that the DT-DOBC of our interest is constructed as in Fig.~\ref{fig:DT-DOB-DT-DOB}, with the DT nominal model $\PPRM_{\mathsf n}^\ddSF(z;\Delta)$ \eqref{eq:DT-DOB-DTNominalModel}, the DT nominal controller $\CCRM^\ddSF(z;\Delta)$ \eqref{eq:DT-DOB-DTController}, and the DT Q-filter $\QQRM^\ddSF(z;\Delta)$ \eqref{eq:DT-DOB-Q}.

\begin{rem}
	As is seen in Fig.~\ref{fig:DT-DOB-DT-DOB_Equiv1}, another widely-used way of realizing the DT-DOB is to redefine the DT nominal model as $\tilde\PPRM^\ddSF_{\mathsf n}(z;\Delta):=z^{q} \PPRM^\ddSF_{\mathsf n}(z;\Delta)$ with a positive integer $q\geq \nu-n_{\mmSF\nnSF}$ (and a redesigned DT Q-filter $\tilde{\QQRM}^\ddSF(z;\Delta)$ whose relative degree can be freely chosen), so that the inverse of the nominal model can stand alone \citep{CT12,Chen2015,TLT00}.
	Even though the two structures in Figs.~\ref{fig:DT-DOB-DT-DOB} and \ref{fig:DT-DOB-DT-DOB_Equiv1} might seem different from each other, they are in fact equivalent in the input-to-output sense under $\QQRM^\ddSF(z;\Delta) = z^{-q}\tilde{\QQRM}^\ddSF(z;\Delta)$. 
	Keeping this equivalence in mind, this paper mainly discusses the DT-DOB structure in Fig.~\ref{fig:DT-DOB-DT-DOB}.
\end{rem}

\begin{figure}
	\begin{center}
		\includegraphics[width=.4\textwidth]{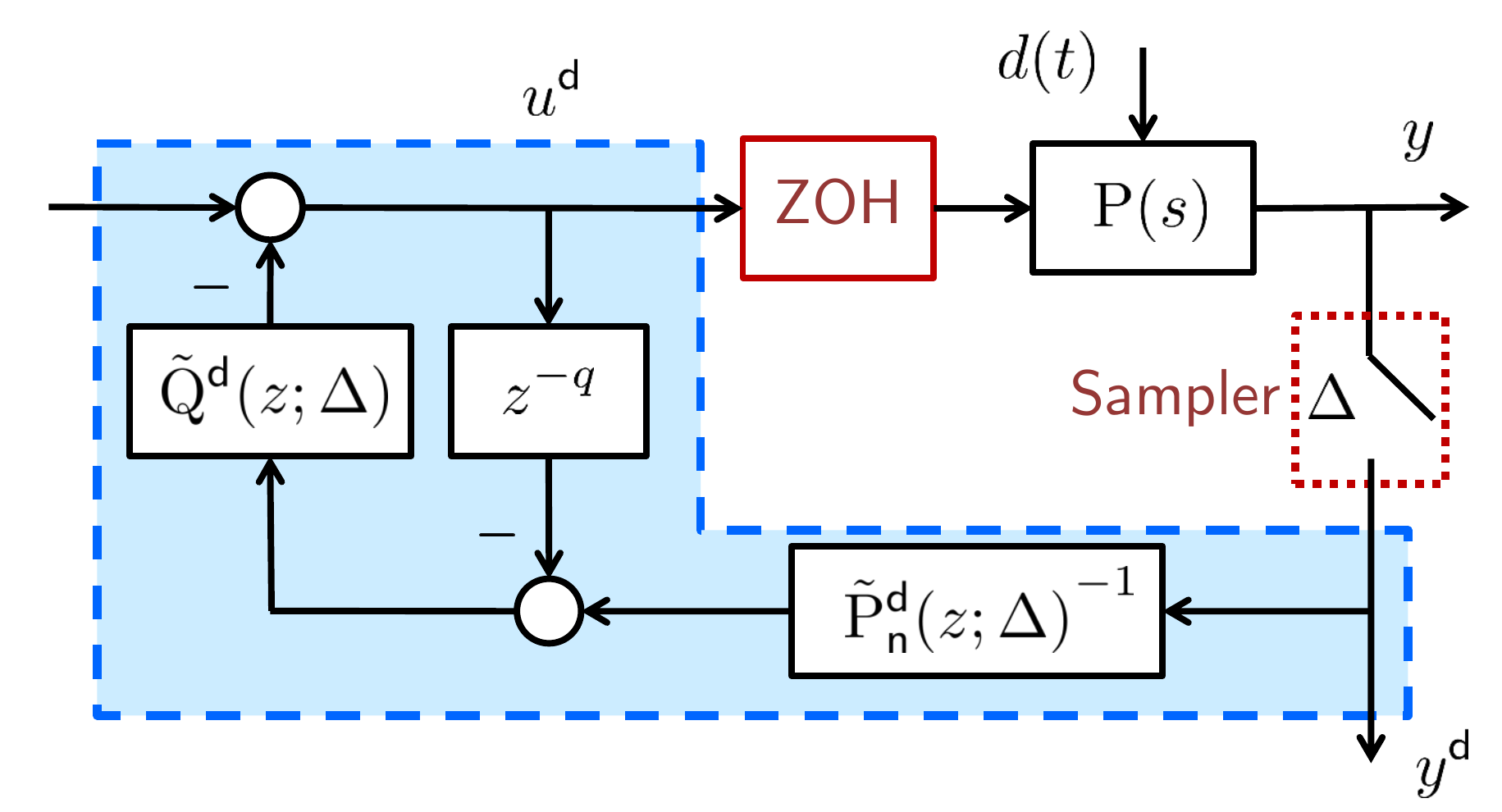}		
	\end{center}
	\caption{Another realization of DT-DOB scheme}\label{fig:DT-DOB-DT-DOB_Equiv1}
\end{figure}

\section[Stability Analysis]{ Robust Stability Condition for Discrete-time Disturbance Observers under Fast Sampling}
\label{sec:DT-DOB-Stability}

Based on the general expression of the DT-DOBC, in this section we present a robust stability condition for  the DT-DOB controlled system in Fig.~\ref{fig:DT-DOB-DT-DOB}, which is the main contribution of this paper. 
To introduce the notion of robust internal stability, define the $\mathcal{Z}$-transform of $e^\ddSF[k]:= r^\ddSF[k] - y^\ddSF[k]$ as $\eeRM^\ddSF(z) := \ZZCAL\{e^\ddSF[k]\}$.
With this symbol, the transfer function matrix from the external inputs $[{\rrRM}^\ddSF(z);\hat{\ddRM}^\ddSF(z)]$ to the internal signals $[\eeRM^\ddSF(z);\uuRM^\ddSF(z);\yyRM^\ddSF(z)]$ is given as follows: (In what follows, we often drop $(z;\Delta)$ or $(z)$ if trivial.)
\begin{align}
& \frac{1}{\QQRM^\ddSF(\PPRM^\ddSF-\PPRM^\ddSF_{\nnSF}) + \PPRM^\ddSF_{\nnSF} (1+\PPRM^\ddSF \CCRM^\ddSF )} \label{eq:DT-DOB-TFM} \\
& \quad\quad\quad \times \begin{bmatrix}
\QQRM^\ddSF(\PPRM^\ddSF-\PPRM^\ddSF_{\mathsf n})+\PPRM^\ddSF_{\mathsf n} & - \PPRM^\ddSF_{\mathsf n}(1-\QQRM^\ddSF) {\WWRM}^\ddSF \\
\PPRM^\ddSF_{\mathsf n} \CCRM^\ddSF & - (\PPRM^\ddSF_{\mathsf n}\CCRM^\ddSF+\QQRM^\ddSF) {\WWRM}^\ddSF\\
\PPRM^\ddSF\PPRM^\ddSF_{\mathsf n} \CCRM^\ddSF & \PPRM^\ddSF_{\mathsf n} (1-\QQRM^\ddSF) {\WWRM}^\ddSF
\end{bmatrix}.\notag
\end{align}
Then the DT-DOB controlled system is said to be {\it internally stable} if all of the transfer functions in \eqref{eq:DT-DOB-TFM} are Schur stable. 
The closed-loop system is said to be {\it robustly internally stable} if it is internally stable for all $\PPRM (s)\in\mathcal{P}$. 
It is then clear that for given $\Delta< \min\{ \Delta^\star_{\mathcal{P}}, \Delta^\star_{\nnSF \ccSF} \}$, the closed-loop system is robustly internally stable if and only if the characteristic polynomial 
\begin{equation}
{\Psi}^\ddSF := \big( \DDRM^\ddSF \DDRM_{\ccSF}^\ddSF + \NNRM^\ddSF \NNRM_{\ccSF}^\ddSF\big) \NNRM_{\nnSF}^\ddSF \DDRM_{\qqSF}^\ddSF + \NNRM_{\qqSF}^\ddSF \DDRM_{\ccSF}^\ddSF \big( \NNRM^\ddSF \DDRM_{\nnSF}^\ddSF - \NNRM_{\nnSF}^\ddSF \DDRM^\ddSF \big) \label{eq:DT-DOB-Psi}
\end{equation}
of \eqref{eq:DT-DOB-TFM} is Schur for all $\PPRM (s)\in\mathcal{P}$.
It is noted that for all $\Delta< \min\{ \Delta^\star_{\mathcal{P}},\Delta^\star_{\nnSF\ccSF}\}$, the degree of $\Psi^\ddSF(z;\Delta)$ is fixed as ${\rm deg}\big({\Psi}^\ddSF(z;\Delta) \big) = n+n_\ccSF + n_\qqSF + (n-\nu+n_{\mmSF\nnSF})=: \overline{n}$.

As aforementioned, the key idea of analyzing the robust internal stability is to investigate the ``limiting'' location of the roots of ${\Psi}^\ddSF(z;\Delta)$ as $\Delta$ is taken sufficiently small.
In the analysis, the following technical lemma will play a crucial role.
\begin{lem}\label{lem:DT-DOB-Approach} 
	Let $\XXRM^\ddSF\in \mathbb{R}[z]$  and $\YYRM^\ddSF \in {\mathbf C}_{\mathbb{R}}[z]$ where $\deg(\XXRM^\ddSF(z))=n$ and $\lim_{T\rightarrow 0^+}\YYRM^\ddSF(z;T) = 0$. 
	Also let $\eta^{\star}_i$, $i=1,\dots,n$, be the roots of $\XXRM^\ddSF(z)$. Then there are $n$ roots of $\XXRM^\ddSF(z)+\YYRM^\ddSF(z;T)$, say $\eta^\ddSF_i(T)$, $i=1,\dots,n$, such that $\lim_{T\rightarrow 0^+}\eta^\ddSF_i(T)=\eta^{\star}_i$ (even if $\XXRM^\ddSF(z)+\YYRM^\ddSF(z;T)$ may have more than $n$ roots for all $T>0$).
\end{lem}

{\bf PROOF.} 
	The lemma is a natural extension of \citet[Lemma 1]{SJ09} and can be proved by the Rouche's theorem \citep{Flanigan83}. 
	We omit its detailed derivation due to page limit.
$\hfill\blacksquare$

As the first step of the analysis, the following lemma shows the limiting location of the roots of ${ \Psi}^\ddSF(z;\Delta)=0$ in the $z$-domain.

\begin{lem}\label{lem:DT-DOB-Fast}
	Suppose that Assumptions~\ref{asm:DT-DOB-Uncertainty} and \ref{asm:DT-DOB-DTC} hold.
	Let $\NNRM_{\mathsf q}^{\star}(z) := \lim_{\Delta\rightarrow 0^+} \NNRM_{\mathsf q}^\ddSF(z;\Delta)$, $\DDRM_{\mathsf q}^{\star}(z) := \lim_{\Delta\rightarrow 0^+} \DDRM_{\mathsf q}^\ddSF (z;\Delta)$, ${h} := 2n - \nu + n_\ccSF$, and 
	\begin{equation}\label{eq:DT-DOB-pf}
	\Psi_{\sf fast}^\ddSF (z) := {\MMRM}_\nnSF^{\star}(z) \big(\DDRM_{\qqSF}^{\star}(z)-\NNRM_{\qqSF}^{\star}(z)\big)+\frac{g}{g_{\nnSF}}{\MMRM}^\star(z) \NNRM^{\star}_{\qqSF}(z).
	\end{equation}
	Then as $\Delta\rightarrow 0^+$,
	$h$ roots of $\Psi^\ddSF(z;\Delta)$ approach $1+j0$, while the remaining 
	$(\overline{n}-h)=(n_{\mmSF\nnSF}+n_\qqSF)$ roots converge to those of $\Psi_{\sf fast}^\ddSF(z)$.
\end{lem}

{\bf PROOF.}
	Without loss of generality, let $\Delta$ be smaller than $\min\{ \Delta^\star_{\mathcal{P}},\Delta^\star_{\nnSF\ccSF}\}$, and define two polynomials $\Psi_1^\ddSF:= \big(\DDRM^\ddSF \DDRM^\ddSF _{\mathsf c} + \NNRM^\ddSF \NNRM^\ddSF_{\mathsf c}\big)\NNRM^\ddSF_{\mathsf n}$ and $\Psi_2^\ddSF:= \big(\NNRM^\ddSF \DDRM^\ddSF _{\mathsf n} - \NNRM^\ddSF_{\mathsf n} \DDRM^\ddSF \big) \DDRM^\ddSF _{\mathsf c}$.	
	Then ${\Psi}^\ddSF(z;\Delta)$ in \eqref{eq:DT-DOB-Psi} is decomposed by ${ \Psi}^\ddSF(z;\Delta) = \Psi_1^\ddSF(z;\Delta){\DDRM^\ddSF _\qqSF}(z;\Delta)+\Psi_2^\ddSF(z;\Delta){\NNRM^\ddSF_\qqSF}(z;\Delta)$. 
	By multiplying $\Delta^{h}$ to these components of the right-hand side, 
	one obtains $\Delta^{h}\Psi_1^\ddSF = (\Delta^n \DDRM^\ddSF ) (\Delta^{n_\ccSF}\DDRM^\ddSF _\ccSF) (\Delta^{n-\nu}\NNRM^\ddSF_\nnSF)  + {\Delta}^{\nu+\nu_\ccSF} (\Delta^{n-\nu}\NNRM^\ddSF) (\Delta^{n_\ccSF - \nu_\ccSF}\NNRM^\ddSF_\ccSF) (\Delta^{n-\nu}\NNRM^\ddSF_\nnSF)$ and ${\Delta}^{h}\Psi_2^\ddSF = (\Delta^{n-\nu} \NNRM^\ddSF) (\Delta^n \DDRM^\ddSF _\nnSF) (\Delta^{n_\ccSF}\DDRM^\ddSF _\ccSF) - (\Delta^{n-\nu}\NNRM^\ddSF_\nnSF) (\Delta^{n} \DDRM^\ddSF ) (\Delta^{n_\ccSF} \DDRM^\ddSF _\ccSF)$.
	From Lemma~\ref{lem:DT-DOB-ZOH} it directly follows that $\Delta^n \DDRM^\ddSF = \prod_{i=1}^{n} \big(z-\ppSF_i^\ddSF(\Delta) \big)$ converges to $(z-1)^n$ as $\Delta\rightarrow 0^+$.
	By repeating similar computations to the components of $\Delta^h \Psi^\ddSF_1$ and $\Delta^h \Psi_2^\ddSF$ (together with Assumption~\ref{asm:DT-DOB-DTC}), we have $\lim_{\Delta \rightarrow 0^+} \Delta^{h}  \Psi_1^\ddSF(z;\Delta) = g_\nnSF {\MMRM}^{\star}_\nnSF(z) (z-1)^{h}$ and $\lim_{\Delta \rightarrow 0^+} \Delta^{h} \Psi_2^\ddSF(z;\Delta) = \big(g \MMRM^{\star}(z)-g_\nnSF {\MMRM}^{\star}_\nnSF(z) \big) (z-1)^{h}$.
	These limits yield
	\begin{align}
	& \lim_{\Delta\rightarrow 0^+} \Delta^{h} { \Psi}^\ddSF (z;\Delta)\notag\\
	& = \lim_{\Delta \rightarrow 0^+} \Delta^{h} \Psi_1^\ddSF(z;\Delta){\DDRM^\ddSF _\qqSF}(z;\Delta) + \Delta^{h}\Psi_2^\ddSF(z;\Delta){\NNRM^\ddSF_\qqSF}(z;\Delta)\notag \\
	& = g_\nnSF (z-1)^{h}  \Big({\MMRM}^{\star}_\nnSF(z)({\DDRM^{\star}_\qqSF}(z) - {\NNRM^{\star}_\qqSF}(z))+\frac{g}{g_\nnSF} \MMRM^{\star} (z) {\NNRM^{\star}_\qqSF} (z)\Big)\notag\\
	& = g_\nnSF (z-1)^{h} \Psi_{\mathsf{fast}}^\ddSF (z)
	\end{align}
	where the degree of $\Psi^\ddSF_{\sf fast}(z)$ is given by $\overline{n} - h$. 
	Noting that the roots of $\Delta^{h} { \Psi}^\ddSF (z;\Delta)$ are the same as those of ${  \Psi}^\ddSF (z;\Delta)$ for any $\Delta>0$, the proof is concluded from Lemma~\ref{lem:DT-DOB-Approach} where $\XXRM^\ddSF(z)$ and $\YYRM^\ddSF(z;\Delta)$ are set as $\XXRM^\ddSF (z)=g_\nnSF(z-1)^{h} \Psi^\ddSF_{\sf fast}(z)$ and $\YYRM^\ddSF(z;\Delta)= \Delta^{h} { \Psi}^\ddSF (z;\Delta) - g_\nnSF (z-1)^{h} \Psi^\ddSF_{\sf fast}(z)$.
$\hfill\blacksquare$

Let us now denote the roots of ${ \Psi}^\ddSF(z;\Delta)$ as $z = \zeta_i^\ddSF(\Delta)$, and rearrange them to satisfy 
$\lim_{\Delta \rightarrow 0^+}\zeta_i^\ddSF(\Delta)=1+j0$ for all $i=1,\dots,h$.    
From the perspective of the singular perturbation theory \citep{KKO99,Litkouhi1985,Yun2017} (especially in the DT domain), we call the first $h$ roots (which converge to $1+j0$) as {\it slow modes} of the DT-DOB controlled system, while the remaining ones as {\it fast modes}.
As seen in Lemma~\ref{lem:DT-DOB-Fast}, for the limiting case (i.e., $\Delta\rightarrow 0^+$) the stability of the fast modes is equivalent to that of the polynomial $\Psi^\ddSF_{\sf fast}(z)$.
On the other hand, whether or not the slow modes are stable is still unanswered, because the lemma tells only that the slow modes converge to $1+j0$ without specifying the direction of the convergence. 
For further discussion, the following lemma is also required.

\begin{lem}\label{lem:DT-DOB-Slow}
	Suppose that Assumptions~\ref{asm:DT-DOB-Uncertainty} and \ref{asm:DT-DOB-DTC} hold.
	Then for all $i=1,\dots,h$,
	the complex variable $\xi^\ddSF_i(\Delta):= (\zeta_i^\ddSF (\Delta)-1)/\Delta$ converges, as $\Delta\rightarrow 0^+$, to the roots   
	$s=\xi_i^\star$ of the polynomial 
	\begin{equation}
	\Psi_{\sf slow}(s) := \NNRM(s) \big( \DDRM_\nnSF(s)\DDRM_\ccSF(s) + \NNRM_\nnSF(s) \NNRM_\ccSF(s)\big)
	\end{equation}
	in which $\deg\big(\Psi_{\sf slow}(s)\big) = 2n-\nu + n_\ccSF = h$.
\end{lem}
{\bf PROOF.} 
	For the proof, we employ a complex variable $\gamma := (z-1)/\Delta$
	which associates with the ``incremental'' operator and approximates the differential operation $s$ of the CT domain \citep{YG14}.
	With this symbol, the characteristic polynomial ${ \Psi}^\ddSF(z;\Delta)$ of our interest is rewritten in the ${\gamma}$-domain as
	\begin{equation}\label{eq:DT-DOB-PsiI}
	{ \Psi}^\iiSF({\gamma};\Delta):={ \Psi}^\ddSF(1+\Delta {\gamma};\Delta).
	\end{equation}	
	The proof is complete by showing that as $\Delta\rightarrow 0^+$, the roots $\gamma = \xi^\ddSF_i(\Delta)$ of ${\Psi}^\iiSF ({\gamma};\Delta)=0$ approach $\xi_i^\star$ for $i=1,\dots,h$. }
		For this, let ${\NNRM^\iiSF}$, $\NNRM^\iiSF_\nnSF$, ${\NNRM^\iiSF_\ccSF}$, ${\NNRM^\iiSF_\qqSF}$, ${\DDRM^\iiSF}$, $\DDRM^\iiSF_\nnSF$, ${\DDRM^\iiSF_\ccSF}$ and ${\DDRM^\iiSF_\qqSF}$ be the polynomials of $\gamma$ corresponding to ${\NNRM^\ddSF}$, $\NNRM^\ddSF_\nnSF$, ${\NNRM^\ddSF_\ccSF}$, ${\NNRM^\ddSF_\qqSF}$, ${\DDRM^\ddSF}$, $\DDRM^\ddSF_\nnSF$, ${\DDRM^\ddSF_\ccSF}$ and ${\DDRM^\ddSF_\qqSF}$, respectively, defined in a similar way of \eqref{eq:DT-DOB-PsiI}. 
		(For instance, $\NNRM^\iiSF(\gamma;\Delta):=\NNRM^\ddSF(1+\Delta\gamma;\Delta)$.)
	Then after some computation, one can readily obtain that
	\begin{align}
	& {\Psi}^\iiSF(\gamma;\Delta) = ( {\DDRM^\iiSF} {\DDRM^\iiSF_{\ccSF}} + {\NNRM^\iiSF} {\NNRM^\iiSF_{\ccSF}}) {\NNRM^\iiSF_{\nnSF}} {\DDRM^\iiSF_\qqSF} +{\NNRM^\iiSF_\qqSF} {\DDRM^\iiSF_{\ccSF}}({\NNRM^\iiSF} {\DDRM^\iiSF_{\nnSF}} - {\NNRM^\iiSF_{\nnSF}} {\DDRM^\iiSF}).\notag
	\end{align}
	Under Assumption \ref{asm:DT-DOB-DTC}, it follows that as $\Delta \rightarrow 0^+$,  ${\NNRM^\iiSF_{\nnSF}}({\gamma};\Delta) \rightarrow \NNRM_{\nnSF}({\gamma})$, ${\DDRM^\iiSF_{\nnSF}} ({\gamma};\Delta) \rightarrow \DDRM_{\nnSF}({\gamma})$, ${\NNRM^\iiSF_{\ccSF}}({\gamma};\Delta) \rightarrow \NNRM_{\ccSF}({\gamma})$, and ${\DDRM^\iiSF_{\ccSF}}({\gamma};\Delta) \rightarrow \DDRM_{\ccSF}({\gamma})$. 
	In addition, by definition one has  ${\NNRM^\iiSF_\qqSF}({\gamma};\Delta) \rightarrow c_0^\star = a_0^\star$, and ${\DDRM^\iiSF_\qqSF}({\gamma};\Delta) \rightarrow a_0^\star$. 
	On the other hand, the DT sampled-data model $\PPRM^\iiSF(\gamma;\Delta):=\NNRM^\iiSF(\gamma;\Delta)/\DDRM^\iiSF(\gamma;\Delta)$ written in the incremental form converges to the corresponding CT plant $\PPRM(\gamma)$ under the fast sampling \citep[Lemma 5.11]{YG14}: to be more specific, as $\Delta\rightarrow 0^+$, ${\NNRM^\iiSF}({\gamma};\Delta) \rightarrow \NNRM ({\gamma})$ and ${\DDRM^\iiSF}({\gamma};\Delta) \rightarrow \DDRM({\gamma})$. 
	Finally, putting all the limits together, we have $\lim_{\Delta \rightarrow 0^+} { \Psi}^\iiSF({\gamma};\Delta)
	 = (\DDRM \DDRM_{\ccSF}+\NNRM \NNRM_{\ccSF})\NNRM_{\nnSF} {a}_0^\star + {a}_0^\star \DDRM_{\ccSF}(\NNRM  \DDRM_{\nnSF}-\NNRM_{\nnSF} \DDRM)  = {a}_0^\star \NNRM (\DDRM \DDRM_{\ccSF}+\NNRM \NNRM_{\ccSF}) = {a}_0^\star  \Psi_{\sf slow}({\gamma})$.
	The remainder of the proof can be derived in a similar way of Lemma~\ref{lem:DT-DOB-Fast}, with the help of Lemma~\ref{lem:DT-DOB-Approach}.
$\hfill\blacksquare$

From the result of Lemma~\ref{lem:DT-DOB-Slow}, it is expected that for sufficiently small $\Delta$, each of the slow modes of the DT-DOB controlled system is approximated by $1 + \Delta \xi_i^\star$, and thus they are located inside the unit circle eventually as long as $\xi_i^\star$ is in the open left-half plane.  
Summing up the observations so far, we present our main result on the stability.

\begin{thm}\label{thm:DT-DOB-Main}
		Suppose that Assumptions \ref{asm:DT-DOB-Uncertainty} and  \ref{asm:DT-DOB-DTC} hold. Then there exists $0<\overline \Delta < \min\{ \Delta^\star_{\mathcal{P}}, \Delta^\star_{\nnSF \ccSF} \}$ such that for all $\Delta\in(0,\overline \Delta)$, the DT-DOB controlled system is robustly internally stable if the following conditions hold:
		\begin{enumerate}[\hspace{3ex}]
			\item[(a)] $\CCRM(s)$ internally stabilizes $\PPRM_{\nnSF}(s)$ (that is, $\NNRM_\nnSF(s)\NNRM_\ccSF(s)+\DDRM_\nnSF(s)\DDRM_\ccSF(s)$ is Hurwitz);
			\item[(b)] $\PPRM (s)\in\mathcal{P}$ is of minimum phase;
			\item[(c)] $\Psi^\ddSF_{\sf fast}(z)$ in \eqref{eq:DT-DOB-pf} is Schur for all $\PPRM (s)\in\mathcal{P}$.
		\end{enumerate}
		Moreover, the converse is also true except the marginal cases (i.e., there is $\overline \Delta_\circ>0$ such that for any $\Delta\in(0,\overline \Delta_\circ)$, the closed-loop system is not robustly internally stable if some roots of $\NNRM_\nnSF(s)\NNRM_\ccSF(s)+\DDRM_\nnSF(s)\DDRM_\ccSF(s)$ or some zeros of $\PPRM (s)$ have positive real part, or some roots of $\Psi^\ddSF_{\sf fast}(z)$ are outside the unit circle).
\end{thm}

{\bf PROOF.} 
The claim follows from Lemmas~\ref{lem:DT-DOB-Fast} and \ref{lem:DT-DOB-Slow}.
In particular, the sufficiency and the necessity of Item (c) for robust stability follow from Lemma \ref{lem:DT-DOB-Slow} with the fact that, given a function $\gamma \in {\mathbf C}_{\mathbb R}$ such that $\lim_{\Delta \to 0^+} \gamma(\Delta) = \gamma^*$ where ${\rm Re}(\gamma^*) < 0$ (or, $>0$), there exists $\Delta_{\gamma} > 0$ such that $\|1+\Delta \gamma(\Delta)\| < 1$ (or, $>1$, respectively) for all $\Delta \in (0,\Delta_{\gamma})$.
And, the sufficiency and the necessity of Items (a) and (b) follow from Lemma \ref{lem:DT-DOB-Fast}.
$\hfill\blacksquare$

We emphasize that, except in the degenerative case when  a root of $\Psi_{\sf slow}(s)$ or $\Psi^\ddSF_{\sf fast}(z)$ lies on the marginally stable region in the CT or the DT domains, the stability condition presented in Theorem~\ref{thm:DT-DOB-Main} is both necessary and sufficient under fast sampling. 
Moreover, our result is obtained based on the general expression of the DT-DOBC in Section~\ref{sec:DT-DOB-DT-DOB} and the CT plants of interest are supposed to have general order and the size of uncertainty is not restricted, from which a wide class of the DT-DOB controlled systems can be dealt with. 

Theorem~\ref{thm:DT-DOB-Main} clarifies the relation between the sampling process, the model uncertainty, the discretization method for the DT nominal model, and the stability of the DT-DOB controlled system.
Among these factors, the sampling process plays a crucial role in determining the stability, because it introduces an ``unstable'' sampling zero whenever the sampling is sufficiently fast and $\nu$ is equal to or larger than 3 (Lemmas~\ref{lem:DT-DOB-ZOH} and \ref{lem:DT-DOB-EF}). 
At first glance, these unstable sampling zeros seem to be a fundamental obstacle for robust stability if one recalls the corresponding robust stability for CT-DOB \citep{SJ07}.
Yet interestingly, in Theorem~\ref{thm:DT-DOB-Main}, there is 
an opportunity to robustly stabilize the DT-DOB controlled system even with the unstable sampling zeros of the DT plant. 
A reasoning for this is that the sampling zeros are sufficiently fast in the sense that they are located away from $1+j0$. 
As a result, they appear in $\Psi^\ddSF_{\sf fast}(z)$ unlike the intrinsic zeros that appear in $\Psi_{\sf slow}(s)$, and therefore, their effect can be compensated by a proper selection of the DT Q-filter.
Indeed, we will show shortly that one can always construct a DT Q-filter (and a DT nominal model) that makes $\Psi_{{\sf fast}}^\ddSF(z)$ always Schur against unstable sampling zeros (as well as model uncertainty of interest). 
Additional remarks on the stability analysis will be presented in Section~\ref{sec:Remark}.

\begin{rem}
	When the discretization method for $\PPRM_\nnSF^\ddSF(z;\Delta)$ and $\CCRM^\ddSF(z;\Delta)$ is particularly chosen as the FDM, the same result of Theorem~\ref{thm:DT-DOB-Main} can be derived in the state space.
	In fact, \citet{Yun2016} presented a state space analysis for the DT-DOB controlled systems by employing the DT singular perturbation theory with a DT sampled-data model obtained by the truncated Tayler series, which decomposes the overall dynamics into slow and fast subsystems under fast sampling.
	It is rather interesting that the stability of each slow and fast subsystem (and thus that of the overall dynamics) is determined by the slow and fast modes presented in this paper, respectively.
	From this aspect, the result of this paper can be viewed as a frequency-domain counterpart of \citet{Yun2016} with an extension: compared with \citet{Yun2016} where only the DT-DOBC obtained via the FDM is considered, the stability analysis in this paper handles a larger class of the DT-DOB controlled systems due to the general expression of the DT-DOBC.
\end{rem}

\section{Further Remarks on Stability Analysis}\label{sec:Remark}

\subsection{Selection of $\PPRM_{\mathsf n}^\ddSF(z;\Delta)$}\label{subsec:DT-DOB-Pn}

Similar to the CT-DOB cases, ${\PPRM^\ddSF_\nnSF}(z;\Delta)$ used in the DT-DOB is usually regarded as a nominal counterpart of the actual sampled-data model  $\PPRM^\ddSF(z;\Delta)$ in the ZOH equivalent form. 
Obviously, even small mismatch between $\PPRM^\ddSF(z;\Delta)$ and $\PPRM^\ddSF_\nnSF(z;\Delta)$ yields additional model uncertainty to be compensated by the DT-DOB.  
From this point of view, the best  candidate for ${\PPRM^\ddSF_\nnSF}(z;\Delta)$ might be the ZOH equivalent model of $\PPRM_\nnSF(s)$, so that $\PPRM(s)=\PPRM_\nnSF(s)$ implies $\PPRM^\ddSF(z;\Delta) = \PPRM_\nnSF^\ddSF(z;\Delta)$. 
However, our stability result shows that such exact discretization of $\PPRM_{\nnSF}(s)$ must be avoided as long as the CT plant $\PPRM(s)$ has high relative degree, even if there is no uncertainty on the CT plant.
(In what follows, we will write ``{\sf ZOH} ({\sf FDM}, {\sf BDM}, {\sf BT}, and {\sf MPZ}, respectively)'' in the subscript of $\PPRM_{\nnSF}^\ddSF(z;\Delta)$ to indicate that $\PPRM_{\mathsf n}^\ddSF(z;\Delta)$ is obtained by discretizing $\PPRM_{\nnSF}(s)$ via the ZOH equivalence method (FDM, BDM, BT, and MPZ, respectively). 
In addition, similar notations will be used for $\CCRM^\ddSF(z;\Delta)$.)

\begin{cor}\label{cor:ZOH}	
	Suppose that Assumptions~\ref{asm:DT-DOB-Uncertainty} and \ref{asm:DT-DOB-DTC} are satisfied with ${\PPRM^\ddSF_\nnSF}(z;\Delta)=\PPRM^\ddSF_{\nnSF,{\sf ZOH}}(z;\Delta)$ and $\nu\geq 3$.
	Then for any $\QQRM^\ddSF(z;\Delta)$ in \eqref{eq:DT-DOB-Q}, there exists $\overline{\Delta}_\circ>0$ such that for any $\Delta\in(0,\overline{\Delta}_\circ)$ and  for any $\PPRM (s)\in {\mathcal P}$, the DT-DOB controlled system is not internally stable.
\end{cor}

{\bf PROOF.}    
	Under the hypothesis, $\Psi_{\sf fast}^\ddSF(z)$ in this case is computed by $\Psi_{\sf fast}^ \ddSF (z)= ({\BBRM_{\nu-1} (z)}/{\nu!}) \big({\DDRM^\star _\qqSF} (z)- {\NNRM^\star_\qqSF} (z)+ (g/g_\nnSF) {\NNRM^\star_\qqSF}(z)\big)$.
	The corollary then directly follows from 
	Lemma \ref{lem:DT-DOB-Fast} and Item (c) of Lemma~\ref{lem:DT-DOB-EF}. 
$\hfill\blacksquare$

Hence in accordance to Corollary~\ref{cor:ZOH}, in the design of the DT-DOB for high-order plants, we recommend using ``approximate'' discretization of $\PPRM_{\nnSF}(s)$, e.g., those listed in Table~\ref{tab:Discretizations}. 
Then, another question arises: which discretization method would be suitable for the DT-DOB designs to guarantee the robust stability? 
As a partial answer to this question, the following proposition suggests to use the FDM or the BDM rather than the BT or the MPZ in discretizing $\PPRM_{\nnSF}(s)$ because the latter two ones may simply violate the stability of the overall system regardless of not only the quantity of model uncertainty but also the bandwidth of the Q-filter.

\begin{prop}\label{prop:Simp}
	Suppose that Assumptions~\ref{asm:DT-DOB-Uncertainty} and \ref{asm:DT-DOB-DTC} hold and the following statements are additionally satisfied:
		\begin{enumerate}[\hspace{3ex}]
		\item[(a)] $\nu = 4l + 3$ or $\nu = 4l + 4$ with a nonnegative integer $l$;
		\item[(b)] $\PPRM_{\nnSF}^\ddSF(z;\Delta) = \PPRM_{\nnSF,{\sf BT}}^\ddSF(z;\Delta)$ or $\PPRM_{\nnSF}^\ddSF(z;\Delta) = \PPRM_{\nnSF,{\sf MPZ}}^\ddSF(z;\Delta)$;
		\item[(c)] The DT Q-filter $\QQRM^\ddSF(z;\Delta)$ is the form 
			\begin{align}\label{eq:DT-DOB-Q1st}
		\QQRM^\ddSF(z;\Delta) = \QQRM^\ddSF_{\sf 1st}(z) :=   \frac{a^\star_0}{(z-1) + a^\star_0}.
		\end{align}
	\end{enumerate}  
	Then for any $a^\star_0>0$, there exists $\overline{\Delta}_\circ>0$ such that for any $\Delta\in(0,\overline{\Delta}_\circ)$ and  for any $\PPRM (s)\in {\mathcal P}$, the DT-DOB controlled system is not internally stable. 
	 $\hfill\square$
\end{prop}

{\bf PROOF.} Note that $\MMRM^\star_{\nnSF}(z)=(z+1)^\nu/2^\nu$ by the assumption.
Then the polynomial $\Psi^\ddSF_{\sf fast}(z)$ is computed by $\Psi^\ddSF_{\sf fast}(z) = {(z-1)(z+1)^\nu}/{2^\nu}  + a_0^\star ({g}/{g_\nnSF}) ({\BBRM_{\nu-1} (z)}/{\nu!})$.
If an open-loop transfer function is given by $\LLRM(z) = {\mathsf K} \BBRM_{\nu-1}(z)/\big((z+1)^\nu (z-1)\big)$ with ${\mathsf K} = (g/g_\nnSF) 2^\nu (1/\nu!) a_0^\star$, then $\Psi^\ddSF_{\sf fast}(z)$ coincides with the characteristic polynomial of the unity feedback system of $\LLRM(z)$.
We now show that the unity feedback system is unstable ``for all'' ${\mathsf K} >0$. 
For simplicity, for now let $\nu = 4l + 3$. 
Then $\LLRM(z)$ has $2l+1$ unstable zeros and $2l+1$ stable zeros, all of which are real and negative (by Lemma~\ref{lem:DT-DOB-EF}).
The largest among the unstable zeros is denoted by $z = \phi<-1$. 
It is then obvious that the number of all the poles and zeros whose real part is larger than $\phi$ is $2l + 1 + \nu + 1 = 6l + 5$ (and thus it is odd), and the smallest among them is the pole at $z = -1$.
Therefore, according to the root locus technique, for each ${\mathsf K}>0$ there always exists a pole of the unity feedback system $\LLRM(z)/(1+\LLRM(z))$ that is real and is located between the pole $z=-1$ of $\LLRM(z)$ and the zero $z = \phi$ of $\LLRM(z)$ (so that it is unstable).
It has been observed so far that $\Psi_{{\sf fast}}^\ddSF(z)$ with $\nu = 4l+3$ is not Schur for any $a^\star_0>0$, while  
one can readily obtain the same result with $\nu = 4l+4$.
From this, the proposition is concluded with Lemma \ref{lem:DT-DOB-Fast}.
$\hfill\blacksquare$

\begin{rem}\label{rem:Simp}
The findings in Corollary~\ref{cor:ZOH} and Proposition~\ref{prop:Simp} partially support the claim of \citet{Kong2013} that it may be beneficial for stability to choose the nominal model $\PPRM_\nnSF^\ddSF(z;\Delta)$ differently from the exact discretization  $\PPRM^\ddSF_{\nnSF,{\sf ZOH}}(z;\Delta)$ (so that mismatch between the DT plant $\PPRM^\ddSF(z;\Delta)$ and its nominal model $\PPRM_{{\mathsf n}}^\ddSF(z;\Delta)$ inevitably appears even if $\PPRM(s) = \PPRM_{{\mathsf n}}(s)$).
In fact, \citet{Kong2013} claim that it is enough for $\PPRM_\nnSF^\ddSF(e^{j\omega\Delta};\Delta)$ to represent $\PPRM^\ddSF_{\nnSF,{\sf ZOH}}(e^{j\omega\Delta};\Delta)$ only for low frequencies where $\QQRM^\ddSF(e^{j\omega\Delta};\Delta) \approx 1$, and they avoid using $\PPRM_\nnSF^\ddSF(e^{j\omega\Delta};\Delta)$ that vanishes around the Nyquist frequency $\omega = \omega_{\sf nyq}$ (which is the case for $\PPRM^\ddSF_{\nnSF, \sf ZOH}(e^{j\omega\Delta};\Delta)$ when $\nu$ is even) to guarantee stability margin under model uncertainty. 
In the same direction, Corollary~\ref{cor:ZOH} and Proposition~\ref{prop:Simp} strengthens this claim by explicitly exhibiting instability caused by the ZOH, the BT, and the MPZ (even without model uncertainty on $\PPRM(s)$), the latter two of which possibly make $\PPRM_\nnSF^\ddSF(e^{j\omega\Delta};\Delta) = 0$ at $\omega_{\sf nyq}$ like the ZOH with even $\nu$.
This phenomenon is intuitively interpreted as a result of employing the inverse model ${\PPRM_\nnSF^{\ddSF}}(z;\Delta)^{-1}$ in the DT-DOB design, which must incur the infinity gain at the Nyquist frequency when the ZOH, the BT, or the MPZ is employed.
\end{rem}

The discussion of this subsection inspires a design guideline for the DT-DOBC to be presented in Section~\ref{sec:DT-DOB-Design}.

\subsection{Bandwidth of $\QQRM^\ddSF(z;\Delta)$}

In the theory of CT-DOB, it has been reported that as long as the CT plant $\PPRM(s)$ is of minimum phase and has no model uncertainty, the CT-DOB controlled system is inherently stable with ``any'' stable Q-filter \citep{CYCKS03}. 
Therefore, it might make sense in some cases to select the CT Q-filter
\begin{align}
\QQRM(s;\tau)  = \frac{c_{m_\qqSF} (\tau s)^{m_\qqSF}+\cdots + c_0}{(\tau s)^{n_\qqSF} + a_{n_\qqSF-1}(\tau s)^{n_\qqSF-1}+\cdots + a_0 } = \frac{\NNRM_\qqSF(s;\tau)}{\DDRM_\qqSF(s;\tau)}\label{eq:DT-DOB-CT-Q}
\end{align} 
as a typical CT low-pass filter with the binomial coefficients
\begin{align}
\QQRM(s;\tau)  = \QQRM_{\sf bin}(s;\tau) := \frac{1}{(\tau s + 1)^{n_{\qqSF}}}\label{eq:DT-DOB-CT-Qbin}
\end{align} 
in which $\tau>0$ is chosen ``arbitrarily'' such that the bandwidth of $\QQRM(s;\tau)$ covers the frequency range of disturbances of interest, with little concern on the stability issue.
However, when it comes to the DT-DOB, this is not the case anymore. 
Such a naive selection of the Q-filter can violate the stability condition for the DT-DOB controlled system, even without any plant uncertainty.
To clarify this point, for now we consider a ``prototypical'' stable DT all-pass filter
\begin{align}
\QQRM^\ddSF(z;\Delta) = \QQRM^\ddSF_{\sf all}(z)=\frac{1}{z^{n_\qqSF}},\label{eq:DT-DOB-AllPassQd}
\end{align} 
which is derived by discretizing the CT Q-filter \eqref{eq:DT-DOB-CT-Qbin} with the binomial coefficients and $\tau = \Delta$,
as the DT Q-filter in the DT-DOB structure.
Then the following proposition indicates that most of the widely-used discretization methods bring negative results on the stability.

\begin{prop}\label{lem:DT-DOB-All-pass}
	Suppose that Assumptions~\ref{asm:DT-DOB-Uncertainty} and \ref{asm:DT-DOB-DTC} hold with one of the following statements satisfied: 
	\begin{enumerate}[\hspace{3ex}]
		\item[(a)] $\nu\geq 3$ and $\PPRM^\ddSF_{\mathsf n}(z;\Delta) = \PPRM^\ddSF_{{\mathsf n},{\sf FDM}}(z;\Delta)$;
		\item[(b)] $\nu\geq 2$ and $\PPRM^\ddSF_{\mathsf n}(z;\Delta) = \PPRM^\ddSF_{{\mathsf n},{\sf BT}}(z;\Delta)$;
		\item[(c)] $\nu\geq 2$ and $\PPRM^\ddSF_{\mathsf n}(z;\Delta) = \PPRM^\ddSF_{{\mathsf n},{\sf MPZ}}(z;\Delta)$.
	\end{enumerate}
	Moreover, assume that $\QQRM^\ddSF(z;\Delta)$ has the form of \eqref{eq:DT-DOB-AllPassQd} with $n_\qqSF \geq \max\{\nu-n_{\mmSF\nnSF},1\}$.  
	Then there exists $\overline{\Delta}_\circ>0$ such that  for any $\Delta\in (0,\overline{\Delta}_\circ)$ and for any $\PPRM (s)\in{\mathcal P}$, the DT-DOB controlled system is not internally stable.
\end{prop}

{\bf PROOF.}
	The proposition is proved by showing that $\Psi_{\sf fast}^\ddSF(z)$ in Theorem~\ref{thm:DT-DOB-Main} is not Schur for all $g/g_\nnSF \in (0,\infty)$. 
	We here provide the proof for Items (b) and (c) only, while the proof for Item (a) can be derived in a similar way.
	Notice that the degree of $\MMRM_\nnSF^\star(z)$ is $\nu$ and thus ${\PPRM^\ddSF_\nnSF}(z;\Delta)$ under consideration is biproper. 
	From this, the degree $n_\qqSF$ of the Q-filter can be arbitrarily set as $n_\qqSF\geq 1$.
	On the other hand, the Q-filter and the nominal model in the considered situation lead to
	\begin{align*}
	\Psi_{\sf fast}^\ddSF(z) &  = \frac{1}{2^\nu} \Big[ (z+1)^\nu (z^{n_\qqSF}-1) \notag\\
	& \quad\quad\quad  + {\sf K} b_{(\nu-1,\nu-1)}z^{\nu-1}+ \cdots + {\sf K} b_{(\nu-1,0)} \Big]
	\end{align*}
	where $b_{(\nu-1,i)}$ are the coefficients of the Euler-Frobenius polynomial ${\BBRM}_{\nu-1}(z)$ and ${\sf K}:=2^\nu (g/g_\nnSF)(1/\nu!)>0$. 
	Now, by applying the Jury's stability test \citep{PN07} to the first two and the last two coefficients of the above polynomial in the bracket and by noting that $b_{(\nu-1,0)}=1$, it follows that $\Psi_{\sf fast}^\ddSF(z)$ above is Schur only if the following two inequalities hold simultaneously: 
	\begin{subequations}\label{eq:DT-DOB-Jury's ineq}
		\begin{align}
		& |{\sf K} -1|<1~\text{and}~ \label{eq:DT-DOB-Jury's ineq 1}\\
		& |({\sf K} -1)^2-1| \label{eq:DT-DOB-Jury's ineq 2}\\
		& \hspace{-1mm} >\begin{cases}
		|(\nu-1)({\sf K} -1)-((1-\nu)+ {\sf K} b_{(\nu-1,1)})|,&\!\!\text{if}~n_\qqSF=1,\\
		|\nu ({\sf K} -1)-( -\nu+ {\sf K} b_{(\nu-1,1)})|, &\!\!\text{if}~n_\qqSF\geq 2.
		\end{cases} \notag
		\end{align}
	\end{subequations}
	Notice that  $b_{(\nu-1,1)}=2^{\nu}-\nu-1$ and $|({\sf K} -1)^2-1|=-({\sf K} -1)^2+1$ for all ${\sf K} $ satisfying $|{\sf K} -1|<1$. 
	Therefore, the inequalities in \eqref{eq:DT-DOB-Jury's ineq} can be rewritten by
	\begin{align*}
	& 0< {\sf K} <2, ~
	- {\sf K} ^2 + 2 {\sf K}  >\begin{cases}
	2^\nu {\sf K},  &\text{if}~n_\qqSF=1,\\
	(2^\nu-1){\sf K}, &\text{if}~n_\qqSF\geq 2.
	\end{cases} 
	\end{align*}
	It is straightforward that under the constraint $0< {\sf K}  <2$, the second inequality is violated in both cases for all $\nu\geq 2$, which completes the proof.
$\hfill\blacksquare$

Roughly speaking, the instability of the overall system seen in Proposition~\ref{lem:DT-DOB-All-pass} can be interpreted as the consequence of employing a DT Q-filter with too large bandwidth for a non-minimum phase DT system (whose unstable zeros are generated by the sampling process). 
With this in mind, a new design method for the DT Q-filter that guarantees robust stability in the presence of unstable sampling zeros will be presented in Section~\ref{sec:DT-DOB-Design}.

\subsection{Indirect Design by Discretizing Continuous-time Disturbance Observer}

A simple way of implementing the DOB scheme in the sampled-data setting would be to discretize a CT-DOB (especially the CT Q-filter in \eqref{eq:DT-DOB-CT-Q}) that is well-designed using the theory of the CT-DOB. 
Yet unfortunately, this ``indirect'' design of DT-DOB is not straightforward in general, because of the effect of the sampling process on the DOB structure. 
As an instance, we already observed in the previous subsection that the DT-DOB controlled system is unstable with the all-pass filter $\QQRM^\ddSF_{\sf all}(z)$.  

Nonetheless, it may be still desired for some designers to obtain a DT-DOB just by discretizing a CT-DOB, without paying too much attention to the stability condition of Theorem~\ref{thm:DT-DOB-Main}. 
In this subsection, we present a way of discretizing a CT-DOB which yields robust internal stability of the overall system in the sense of Theorem~\ref{thm:DT-DOB-Main}. 
Motivated by the discussions on Proposition~\ref{lem:DT-DOB-All-pass}, the basic idea for this indirect design is to restrict the bandwidth of the CT Q-filter \eqref{eq:DT-DOB-CT-Q} much below the Nyquist frequency.  
Another point we should focus on is that some of the usual discretization methods such as the BT or the MPZ possibly forces the system to be unstable (as seen in Proposition~\ref{prop:Simp}). 
Putting these together, we take $\tau$ in the CT Q-filter as $\tau = \psi \Delta$ with a large constant $\psi>1$, and  discretize the CT-DOB-based controller consisting of $\PPRM_\nnSF(s)$, $\CCRM(s)$, and $\QQRM(s;\tau)$ (in \eqref{eq:DT-DOB-CT-Q}) using the FDM. 
The resulting DT-DOBC is composed of the DT nominal model $\PPRM^\ddSF_{\nnSF}(z;\Delta) = \PPRM^\ddSF_{\nnSF,{\sf FDM}}(z;\Delta)$, the DT nominal controller $\CCRM^\ddSF(z;\Delta) = \CCRM^\ddSF_{{\sf FDM}}(z;\Delta)$, and the DT Q-filter
\begin{align}
& \QQRM^\ddSF(z;\Delta) = \QQRM^\ddSF_{\sf ind}(z;\Delta)\label{eq:DT-DOB-Indirect}\\
& \quad  := \frac{c_{m_\qqSF} \big( \psi (z-1)\big)^{m_\qqSF}+\cdots + c_0}{\big(\psi (z-1)\big)^{n_\qqSF} + a_{n_\qqSF-1}\big(\psi (z-1)\big)^{n_\qqSF-1}+\cdots + a_0 }\notag
\end{align}
where $\psi=\tau/\Delta$ denotes the ratio between $\tau$ and $\Delta$.
Then the following proposition reveals that as long as the cutoff frequency of the CT Q-filter is far below the Nyquist frequency of the sampled-data system, the robust stability condition of the DT-DOB controlled system in Theorem~\ref{thm:DT-DOB-Main} is ``relaxed'' to that of the CT-DOB controlled system with $\QQRM(s;\tau)$ presented in the previous works of \citet{SJ09,SJ07} on the CT-DOB. 
\begin{prop}\label{prop:DT-DOB-Indirect}
	Suppose that Assumptions~\ref{asm:DT-DOB-Uncertainty} and \ref{asm:DT-DOB-DTC} hold with $\PPRM^\ddSF_\nnSF(z;\Delta)=\PPRM^\ddSF_{\nnSF,{\sf FDM}}(z;\Delta)$, $\CCRM^\ddSF(z;\Delta) = \CCRM^\ddSF_{\sf FDM}(z;\Delta)$, and $\QQRM^\ddSF(z;\Delta) = \QQRM^\ddSF_{\sf ind}(z;\Delta)$ in \eqref{eq:DT-DOB-Indirect} whose coefficients $a_i$ and $c_i$ are selected such that\footnote{For the design method of the CT Q-filter $\QQRM(s;\tau)$ such that \eqref{eq:DT-DOB-FastCT} becomes Hurwitz for all $\PPRM\in \mathcal{P}$, the readers are referred to \citet{SJ09}.} 
	\begin{align}\label{eq:DT-DOB-FastCT}
	\Psi_{\sf fast,ind}(s) := \DDRM_\qqSF(s;1) - \NNRM_\qqSF(s;1) + \frac{g}{g_\nnSF} \NNRM_\qqSF(s;1)
	\end{align}
	is Hurwitz for all $\PPRM (s)\in\mathcal{P}$.
	Then there is $\underline{\psi}>0$ such that for each $\psi>\underline{\psi}$, the polynomial $\Psi_{{\sf fast}}^\ddSF(z)$ in \eqref{eq:DT-DOB-pf} is Schur for all $\PPRM (s)\in\mathcal{P}$.
	Moreover, if Items (a) and (b) of Theorem~\ref{thm:DT-DOB-Main} additionally hold, then for each $\psi>\underline{\psi}$, there exists $0<\overline \Delta = \overline{\Delta}(\psi) < \min\{ \Delta^\star_{\mathcal{P}}, \Delta^\star_{\nnSF \ccSF} \}$ such that for all $\Delta \in (0,\overline{\Delta})$, the DT-DOB controlled system is robustly internally stable.
\end{prop}

{\bf PROOF.}
	For the proof of the first part, we note that $\DDRM_\qqSF^\star(z) = \DDRM_\qqSF(\psi(z-1);1)$ and $\NNRM_\qqSF^\star(z) = \NNRM_\qqSF(\psi(z-1);1)$. 
	Thus one has $\Psi^\ddSF_{\sf fast}(z) = \big(\DDRM_\qqSF(\psi(z-1);1)-	\NNRM_\qqSF(\psi(z-1);1)\big) + ({g}/{g_{\nnSF}}){\MMRM}^\star(z) \NNRM_\qqSF(\psi(z-1);1)$.
	Now, define a new complex variable $\Gamma:= \psi(z-1)$ (so that $z = 1 + \Gamma/\psi$).
	Then $\Psi^\ddSF_{\sf fast}(z)$ is rewritten in the $\Gamma$-domain by 
	\begin{align*}
	& \hat\Psi_{\sf fast,ind}(\Gamma;\psi) := \Psi_{\sf fast}^\ddSF(1+\Gamma/\psi)\\
	& \quad  = \big(\DDRM_\qqSF(\Gamma;1)-					\NNRM_\qqSF(\Gamma;1)\big)  + \frac{g}{g_{\nnSF}}{\MMRM}^\star(1+\Gamma/\psi) \NNRM_\qqSF(\Gamma;1).
	\end{align*}
	Since $\deg(\MMRM^\star) = \nu-1$, two polynomials $\hat\Psi_{\sf fast,ind}(\Gamma;\psi)$ and $\Psi_{\sf fast,ind}(s)$ have the same number of roots.
	Then, with $\XXRM^\ddSF(\Gamma) = \Psi_{\sf fast,ind}(\Gamma)$ and $\YYRM^\ddSF(\Gamma;1/\psi) = \hat \Psi_{\sf fast,ind}(\Gamma;\psi) - \Psi_{\sf fast,ind}(\Gamma)$, Lemma~\ref{lem:DT-DOB-Approach} guarantees, by $\MMRM^\star(1)=1$, that all the roots of $\hat\Psi_{\sf fast,ind}(\Gamma;\psi)$ converges to the roots of ${\Psi}_{\sf fast,ind}(\Gamma)$ as $\psi \to \infty$.
	The remaining proof proceeds similarly to that of Theorem~\ref{thm:DT-DOB-Main}.
$\hfill\blacksquare$

Proposition~\ref{prop:DT-DOB-Indirect} introduces an alternative way of designing the DT-DOBC from the CT-DOB theory, without much concern on the sampling process. 
Nonetheless, constructing a DT-DOBC in the DT domain directly at the cost of complexity (as in the next section) is worth pursuing.
For instance, we will see shortly in the simulation part that the direct design method can admit a larger bandwidth of the DT Q-filter, by which the disturbance rejection ability of the DT-DOB becomes significantly improved compared with the indirect design of the DT-DOBC based on the result of Proposition~\ref{prop:DT-DOB-Indirect}.

\section[Design Methods]{Systematic Design Guideline for Discrete-time Disturbance Observers-based Controller}\label{sec:DT-DOB-Design}	

In this section, we present a systematic design procedure for the DT-DOBC that achieves robust stabilization for minimum phase CT plants \eqref{eq:DT-DOB-Plant} under arbitrarily large (but bounded) parametric uncertainty stated in Assumption~\ref{asm:DT-DOB-Uncertainty} and sufficiently fast sampling.    
First, select the CT nominal model $\PPRM_\nnSF(s)$ and the CT nominal controller $\CCRM(s)$ such that the CT nominal closed-loop system is internally stable.
Then motivated by the discussions in Subsection~\ref{subsec:DT-DOB-Pn}, we discretize the CT transfer functions such that Assumption~\ref{asm:DT-DOB-DTC} holds and the polynomial $\MMRM^{\star}_\nnSF(z)$ in the numerator of ${\PPRM^\ddSF_\nnSF}(z;\Delta)$ is Schur (e.g., the FDM or the BDM).

For ease of construction,  we set the DT Q-filter $\QQRM^\ddSF(z;\Delta)$ as the DT low-pass filter whose numerator is constant; i.e., 
\begin{align}
& \QQRM^\ddSF(z;\Delta) = \QQRM^\ddSF_{\sf prop}(z;\Delta) \label{eq:DT-DOB-SimpleQ} \\
& := \frac{{a}_{{\sf prop},0}^\star}{(z-1)^{n_\qqSF}+{a}_{{\sf prop},{n_\qqSF}-1}^\star (z-1)^{{n_\qqSF}-1}+\cdots+{ a}_{{\sf prop},0}^\star} \notag
\end{align}
where ${a}_{{\sf prop},i}^\star$, $i=0,\dots,n_\qqSF-1$, are design parameters to be determined below, and $n_\qqSF$ can be arbitrarily chosen as an integer larger than $\max\{ \nu - n_{\mmSF\nnSF},1 \}$.
Notice that the polynomial $\Psi_{\sf fast}^\ddSF(z)$ in \eqref{eq:DT-DOB-pf} turns out to be
\begin{align*}
& \Psi_{{\sf fast}}^\ddSF(z)  = {\MMRM}_\nnSF^\star (z) (z-1) \VVRM(z)  +\frac{g}{g_\nnSF}{\MMRM}^\star(z) {a}_{{\sf prop},0}^\star
\end{align*}
in which $\VVRM(z):=(z-1)^{n_\qqSF-1}+{a}_{{\sf prop},n_\qqSF-1}^\star(z-1)^{n_\qqSF-2}+\cdots+{a}_{{\sf prop},1}^\star$.
The coefficients $a_{{\sf prop},i}^\star$ of the Q-filter are selected as the following steps. 
First, take ${a}_{{\sf prop},1}^\star,\dots,{a}_{{\sf prop},{n_\qqSF-1}}^\star$ to make $\VVRM(z)$ above
Schur. 
Here it is noticed that by Item~(d) of Lemma~\ref{lem:DT-DOB-EF}, 
all the zeros of the open-loop transfer function $\LLRM(z) = {\mathsf K} {\MMRM^\star(z)}/\big({\MMRM_\nnSF^\star(z) (z-1) \VVRM(z)}\big)$ with $\KKSF:= (g/g_\nnSF) a_{\sf prop,0}^\star$  have the real part smaller than $0$, while the poles of $\LLRM(z)$ except one at the marginal point $z = 1+j0$ are all Schur.
Therefore, by applying the root locus technique to $\LLRM(z)$, one can find a sufficiently small $\overline{\KKSF}>0$ such that, for all $0< \KKSF < \overline{\KKSF}$, the polynomial $ {\MMRM}_\nnSF^\star (z) (z-1) \VVRM(z)  + \KKSF {\MMRM}^\star(z)$ is Schur.
Finally, choose $a_{{\sf prop},0}^\star \in \big(0, (g_{\nnSF}/ \overline{g})(1/\overline\KKSF)\big)$ so that  $\overline\KKSF > (\overline{g}/g_\nnSF) a_{{\sf prop},0}^\star \geq (g/g_\nnSF) a_{{\sf prop},0}^\star >0$ for all $g\in[\underline{g},\overline{g}]$. 
By definition, this implies that $\Psi_{{\sf fast}}^\ddSF(z)$ is Schur for all $g\in[\underline{g},\overline{g}]$.

We summarize the contents of this subsection as follows.
\begin{prop}
	Suppose that Assumption~\ref{asm:DT-DOB-Uncertainty} and Item (b) of Theorem~\ref{thm:DT-DOB-Main} hold. 
	Then there exists $0 < \overline \Delta < \min\{ \Delta^\star_{\mathcal{P}}, \Delta^\star_{\nnSF \ccSF} \}$ such that for all $\Delta \in (0,\overline{\Delta})$, the DT-DOB controlled system with the DT-DOBC obtained by the proposed design guideline is robustly internally stable. 
\end{prop}

\section{Simulation Results: Two-mass-spring System}\label{sec:DT-DOB-Sim}

In order to verify the validity of our theoretical results, we perform simulations for two-mass-spring systems in the benchmark problem \citep{WB92,BHLO06}.
The considered CT plant here is the 4th-order two-mass-spring system $\PPRM(s) = K/(M_1 M_2 s^4 + K(M_1 + M_2)s^2)$ 
where $M_1\in [0.5,2]$ and $M_2 \in [0.5,2]$ are the uncertain masses of the carts, and $K\in [0.8,1.2]$ is the unknown spring coefficient.
As a nominal counterpart, $\PPRM_\nnSF(s) = K_\nnSF/(M_{\nnSF,1} M_{\nnSF,2} s^4 + K_\nnSF(M_{\nnSF,1} + M_{\nnSF,2})s^2)$ with the nominal parameters $M_{\nnSF,1} = M_{\nnSF,2} = 1$ and $K_{\nnSF}=1$. The CT nominal controller ${\CCRM}(s) = (-6.83 s^2 + 1.85 s + 0.28)/(s^2 + 4.28 s + 6.08)$ is designed to stabilize the CT nominal model \citep{BHLO06}. 

\begin{figure}[!t]
	\begin{center}
		{
			\subfigure[CT-DOB controlled systems]	{\includegraphics[width=0.34\textwidth]{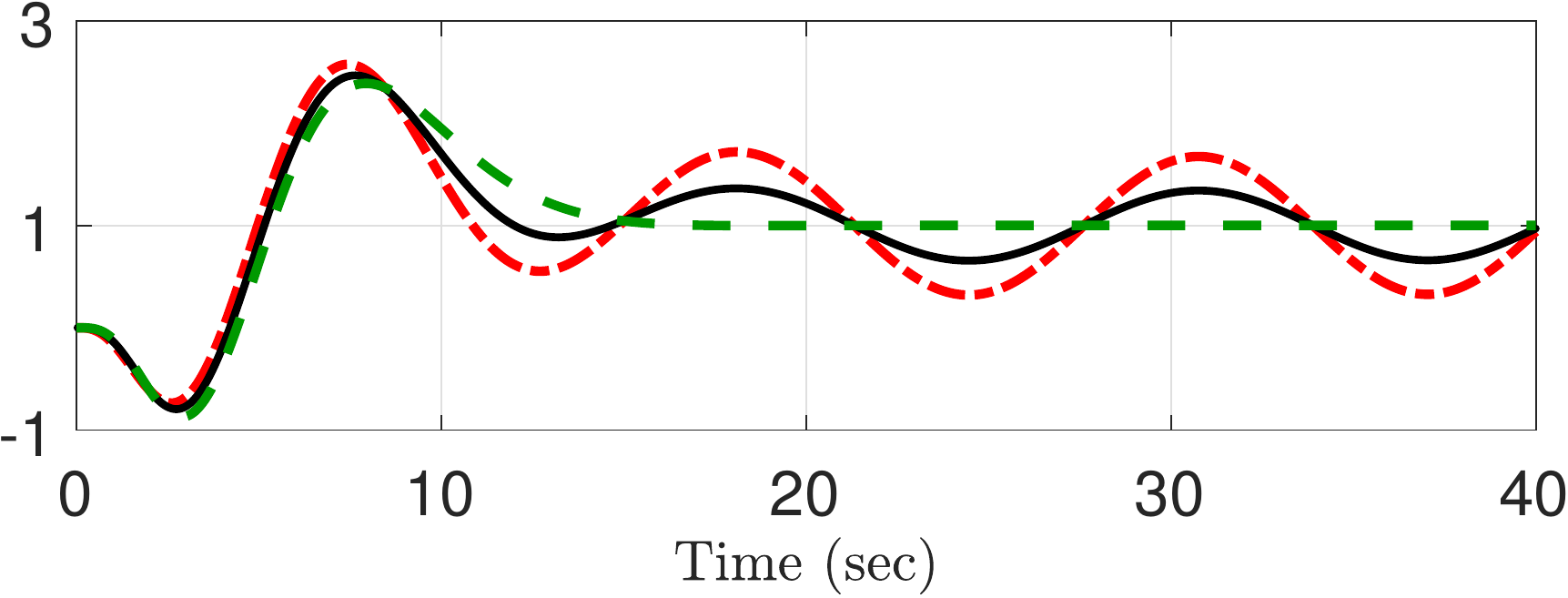}}	\\			
			\subfigure[DT-DOB controlled systems (Enlargement)]	{\includegraphics[width=0.34\textwidth]{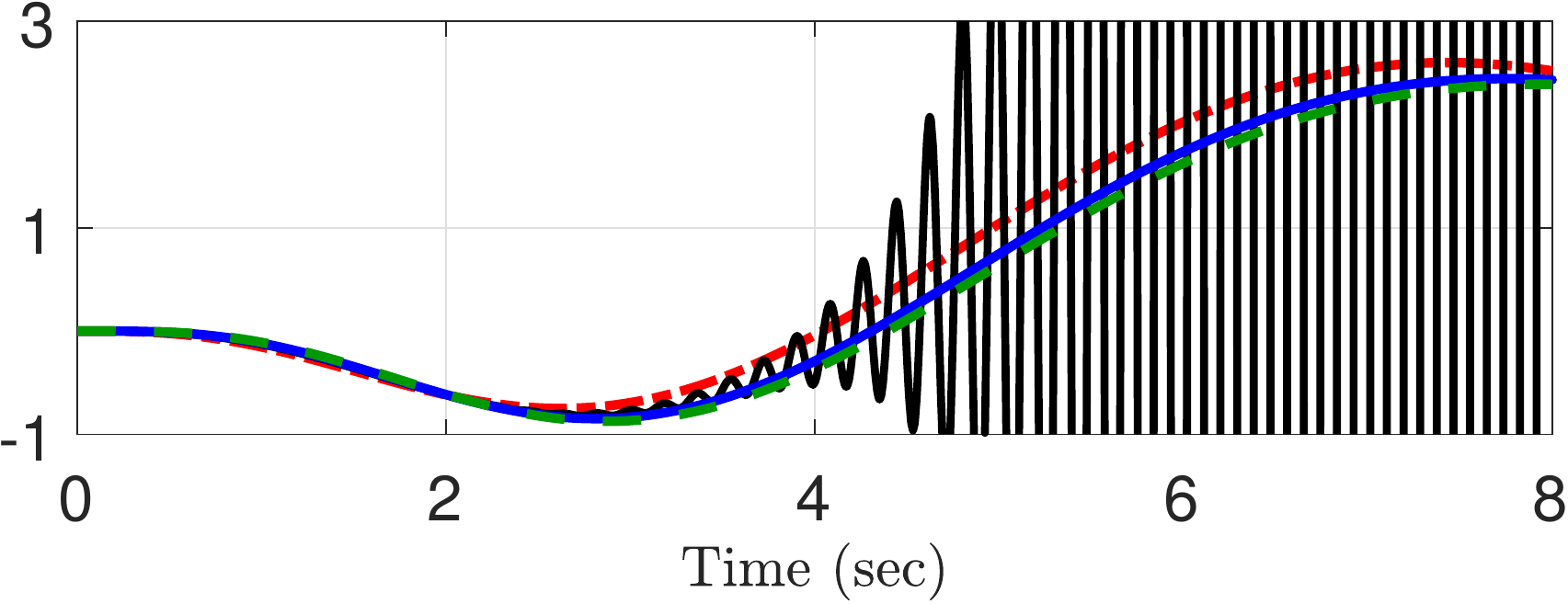}}\\
			\subfigure[DT-DOB controlled systems]	{\includegraphics[width=0.34\textwidth]{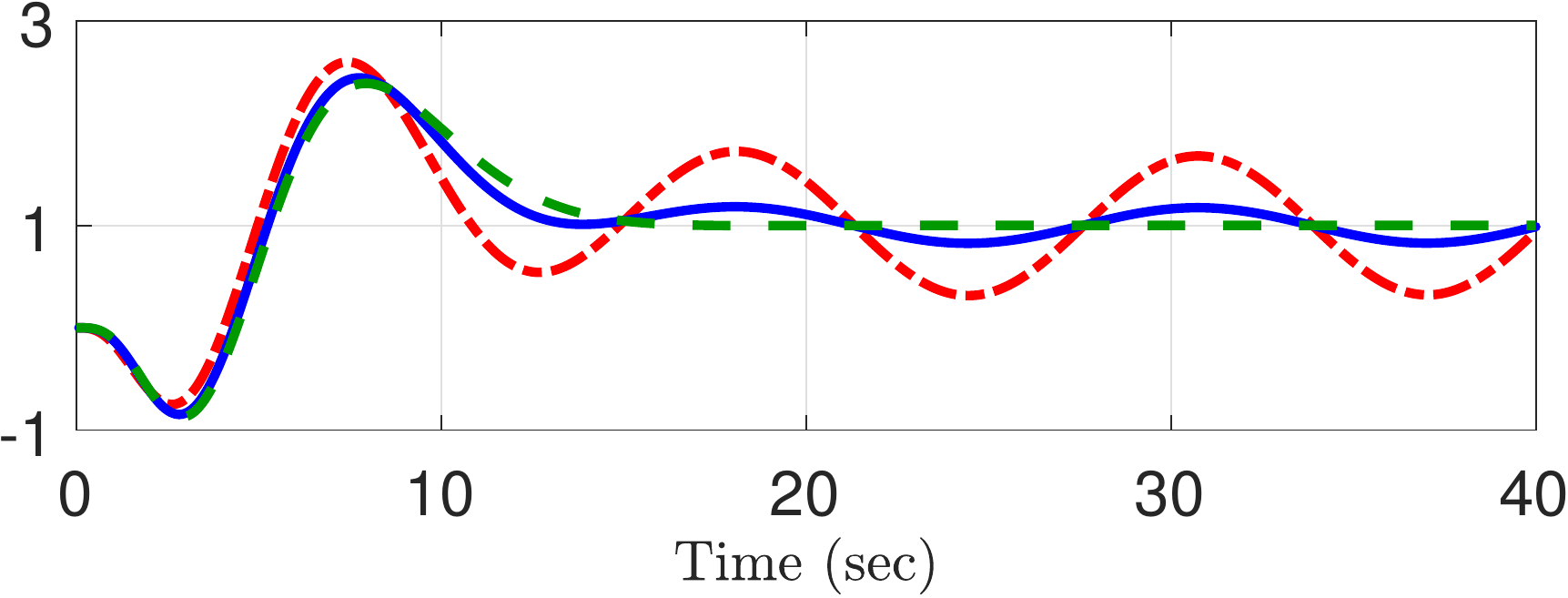}}
		}
	\end{center}
	\caption{Step response $y(t)$ with different designs of Q-filters: CT nominal closed-loop system (dashed green); CT-DOB controlled systems with $\tau = 0.05$ (dash-dotted red) and $\tau = 0.025$ (solid black); and DT-DOB controlled systems  with $\QQRM^\ddSF_{\sf prop}(z;\Delta)$ (solid blue), $\QQRM^\ddSF_{\sf ind, LBW}(z;\Delta)$ (solid black), and $\QQRM^\ddSF_{\sf ind, SBW}(z;\Delta)$ (dash-dotted red)}\label{fig:DT-DOB-SIM1}
\end{figure}

\begin{figure}[!t]
	\begin{center}
		{
			\subfigure[Bode plot of $\QQRM^\ddSF(z;\Delta)$]	{\includegraphics[width=0.4\textwidth]{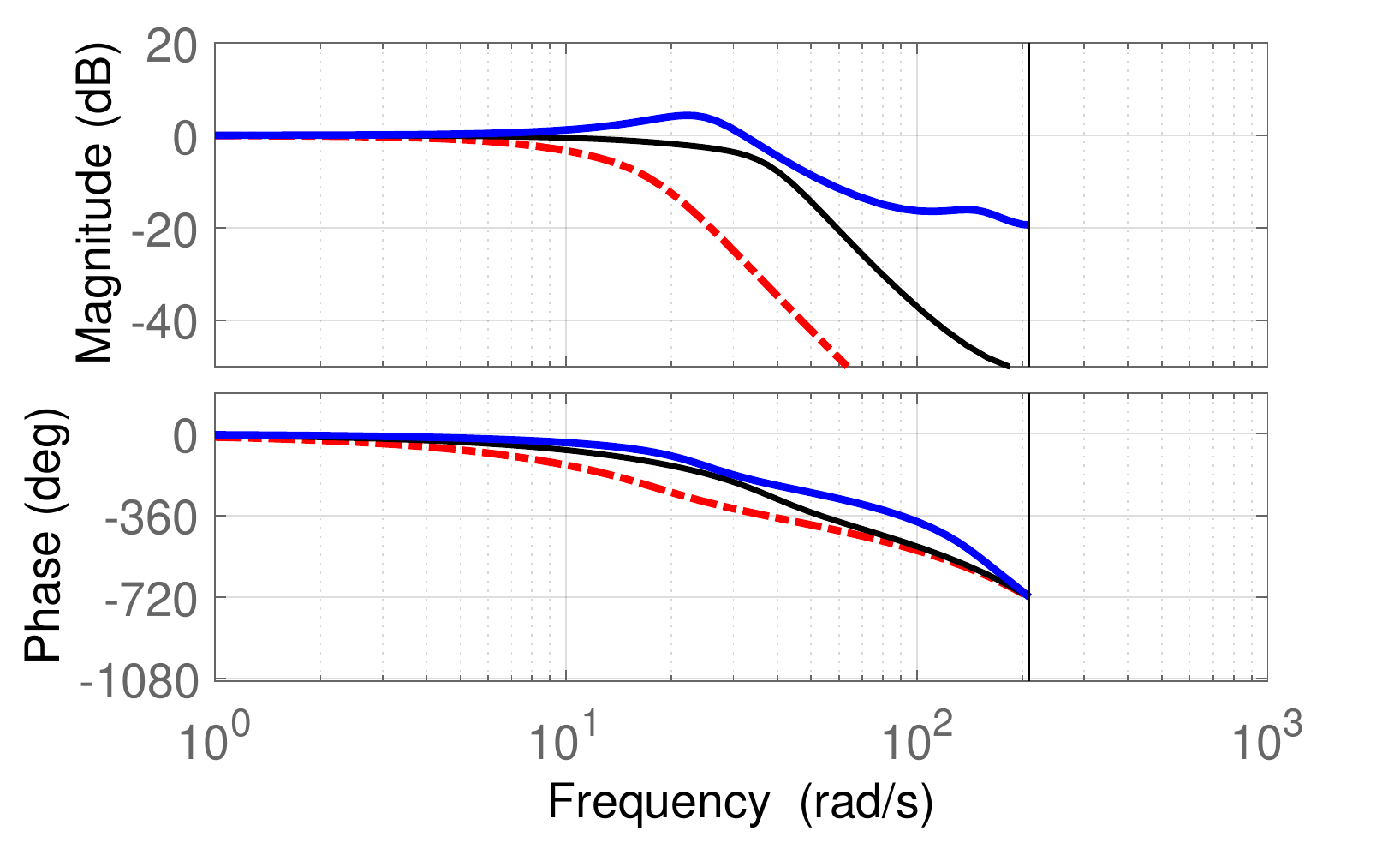}}			
			\subfigure[Sensitivity function]	{\includegraphics[width=0.4\textwidth]{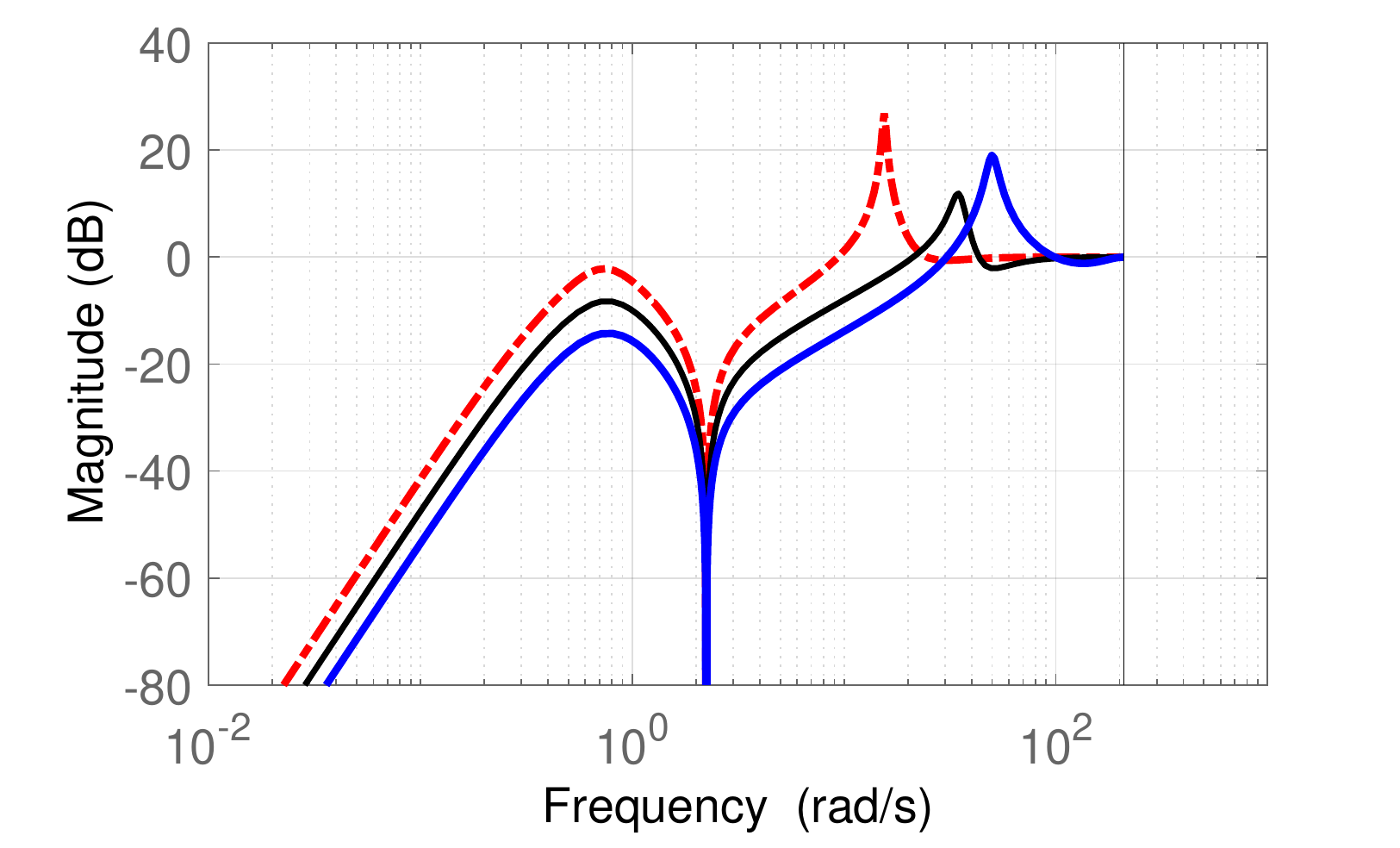}}
		}
	\end{center}
	\caption{Frequency responses of DT Q-filters and sensitivity functions of DT-DOB controlled systems with $\QQRM^\ddSF_{\sf prop}(z;\Delta)$ (solid blue), $\QQRM^\ddSF_{\sf ind, LBW}(z;\Delta)$ (solid black), and $\QQRM^\ddSF_{\sf ind, SBW}(z;\Delta)$ (dash-dotted red)}\label{fig:DT-DOB-Sensitivity}
\end{figure}

\subsection{Simulation 1: Direct vs. Indirect DT-DOB Designs}

The main purpose of this subsection is to compare the DT-DOB obtained via the proposed design guideline with those via discretization of a CT-DOB.
In both cases, the DT nominal model $\PPRM_{\mathsf n}^\ddSF(z;\Delta)$ and the DT nominal controller $\CCRM^\ddSF(z;\Delta)$ are set as $\PPRM_{\mathsf n,{\sf FDM}}^\ddSF(z;\Delta)$ and $\CCRM^\ddSF_{\sf FDM}(z;\Delta)$ using the FDM, respectively.
Now, we construct three types of DT Q-filters as follows. 
The first DT Q-filter is derived by the proposed design guideline in Section~\ref{sec:DT-DOB-Design} as 
$\QQRM^\ddSF_{\sf prop}(z;\Delta) = {a_{{\sf prop},0}^\star}/\big({(z-1)^4 + a_{{\sf prop},3}^\star(z-1) + \cdots + a_{{\sf prop},0}^\star}\big)$
with $[a_{{\sf prop},3}^\star,\dots,a_{{\sf prop},0}^\star] = [3,3,1,0.24]$ so that the corresponding $\Psi_{{\sf fast}}^\ddSF(z)$ in \eqref{eq:DT-DOB-pf} is Schur for all $\PPRM(s)\in \mathcal{P}$.
On the other hand, the remaining two DT Q-filters are derived from their CT counterpart $\QQRM(s;\tau) = {a_0}/\big({(\tau s)^4 + a_3 (\tau s)^3 + a_2 (\tau s)^2 + a_1 (\tau s) + a_0}\big)$
with $[a_3,\dots,a_0] = [2,2,1,0.3]$. 
Note that $a_i$ in the CT Q-filter satisfy the stability condition presented in the theory of CT-DOB (i.e., $\Psi_{{\sf fast,ind}}(s)$ in \eqref{eq:DT-DOB-FastCT} is Hurwitz), and thus the corresponding CT-DOB controlled system is robustly stable as long as $\tau$ is sufficiently small. 
Now, to implement the CT Q-filter in discrete time, we discretize $\QQRM(s;\tau)$ above via the FDM, particularly with two different values of $\tau$, as $\QQRM_{\sf ind,LBW}^\ddSF(z;\Delta) := \QQRM \big((z-1)/\Delta;0.025\big)$ and $\QQRM_{\sf ind,SBW}^\ddSF(z;\Delta) := \QQRM \big((z-1)/{\Delta};0.05\big)$.
We remark that by construction, the bandwidth of $\QQRM_{\sf ind,LBW}^\ddSF$ (with smaller $\tau$) is larger than that of $\QQRM_{\sf ind,SBW}^\ddSF$. 

For time-domain simulations, let $\Delta = 0.015$ and set the uncertain parameters and disturbance of the CT plant as $M_1 = M_2 = 0.8$, $K = 2$, and $d(t) = 0.5 \sin t$.
The simulation results with three different DT Q-filters are depicted in Fig.~\ref{fig:DT-DOB-SIM1}.
Notice that even though the CT-DOBs with both $\tau=0.025$ and $\tau = 0.05$ guarantee robust stability in the CT domain, their discretization leads to different consequences in the DT domain in the end.
This is mainly because the DT Q-filter $\QQRM_{\sf ind,SBW}^\ddSF(z;\Delta)$ associated with larger $\tau = 0.05$ makes the corresponding $\Psi_{{\sf fast}}^\ddSF(z)$ to be Schur, whereas the stability of the polynomial is immediately lost when smaller $\tau = 0.025$ is used. 
On the other hand, the DT-DOB obtained by the proposed design guideline robustly stabilizes the closed-loop system. 
We point out that as seen in Figs.~\ref{fig:DT-DOB-SIM1} and \ref{fig:DT-DOB-Sensitivity}, the proposed DT Q-filter $\QQRM^\ddSF_{\sf prop}(z;\Delta)$ has much larger bandwidth than the DT Q-filter $\QQRM_{\sf ind,SBW}^\ddSF(z;\Delta)$ obtained from the indirect design method, by which the DT-DOB with the former Q-filter shows better disturbance rejection performance in a wide frequency range.

\begin{figure}[!t]
	\begin{center}
		{
			\subfigure[Roots $z = \zeta_i^\ddSF(\Delta)$ of $\Psi^\ddSF(z;\Delta)=0$ under variation of $\Delta$ (blue line) and those of $\Psi^\ddSF_{\sf fast}(z)$ (red marked) (in $z$-domain)]	{\includegraphics[width=0.38\textwidth]{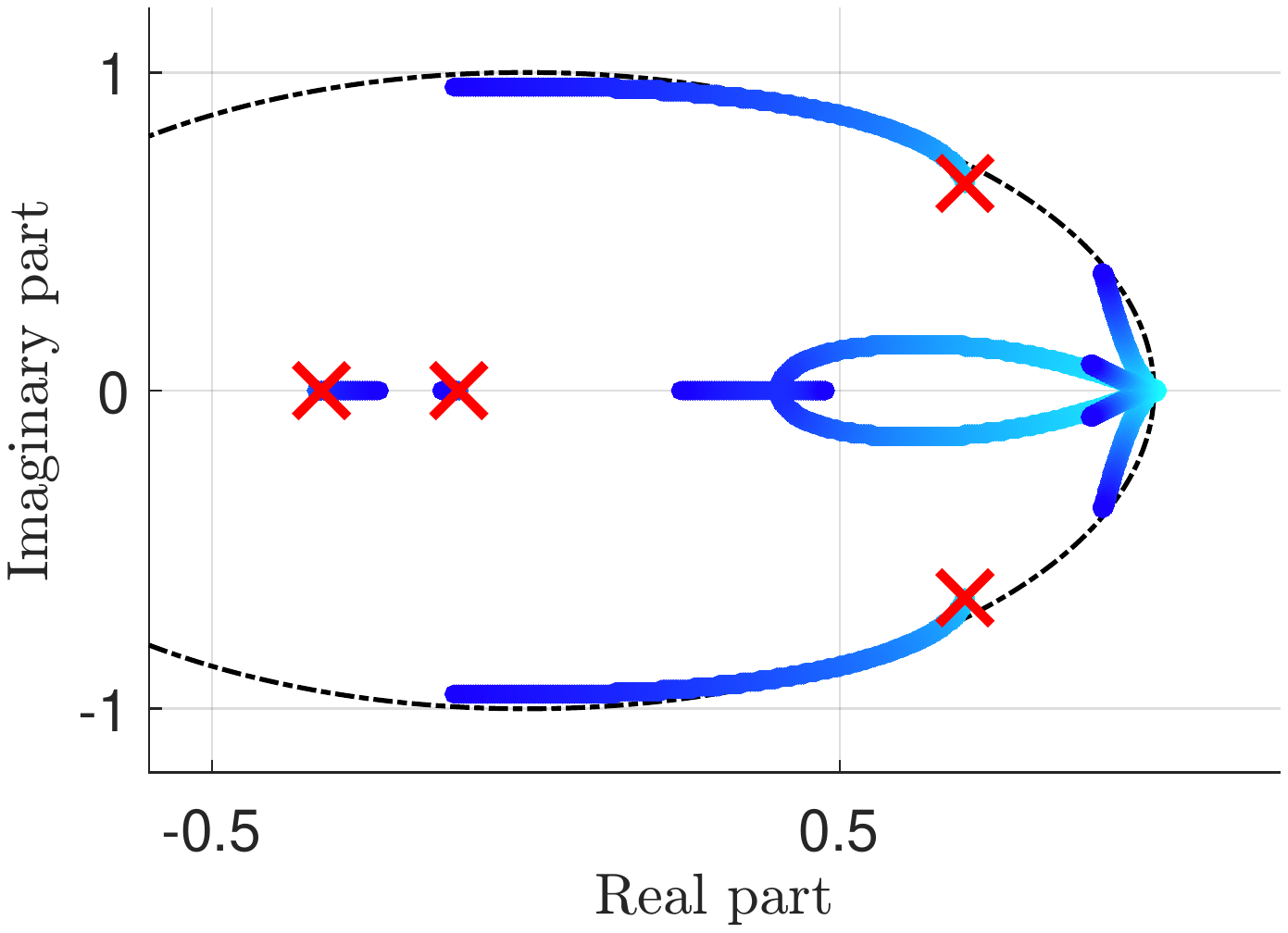}}	
			\subfigure[Roots $\gamma = 1 + \Delta \zeta_i^\ddSF(\Delta)$ of ${ \Psi}^\iiSF({\gamma};\Delta)$ under variation of $\Delta$ (blue line) and those of $\Psi_{\sf slow}(\gamma)$ (green marked) (in $\gamma$-domain)]	{\includegraphics[width=0.38\textwidth]{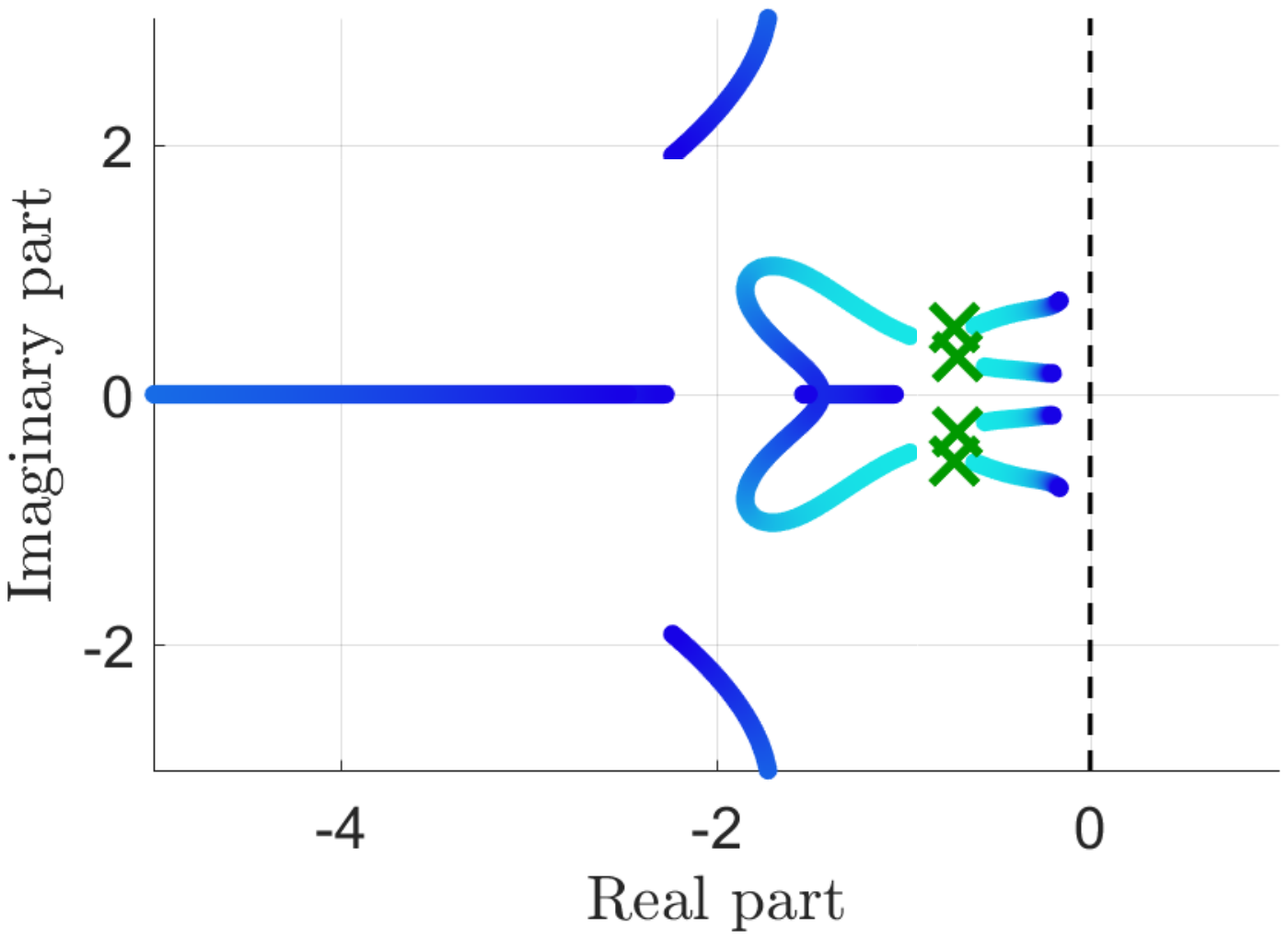}}	
		}
	\end{center}
	\caption{Root contours of characteristic polynomial of DT-DOB controlled system under variation of $\Delta\in [0.001,0.3]$: The smaller $\Delta$ becomes, the brighter the color is.}\label{fig:DT-DOB-RootContour}
\end{figure}

In order to investigate the limiting behavior with respect to the variation of $\Delta$, in Fig.~\ref{fig:DT-DOB-RootContour} we additionally depict root contours of the characteristic polynomial $\Psi^\ddSF(z;\Delta)$ of the DT-DOB controlled system associated with the proposed Q-filter $\QQRM^\ddSF(z;\Delta) = \QQRM^\ddSF_{\sf prop}(z;\Delta)$, with respect to the variation on the sampling period $\Delta\in [0.001,0.3]$. 
It is shown that the four roots of $\Psi^\ddSF(z;\Delta)$ converge to the roots of $\Psi_{{\sf fast}}^\ddSF(z)$ (as the fast modes) in the $z$-domain, all of which are located in the stable region. 
On the other hand, the remaining six roots approach the marginal point $z  = 1 + j0$ from the inside of the unit circle, because the complex variables $\gamma = \xi^\ddSF(\Delta)$ go to the roots of $\Psi_{\sf slow}(s)$ in the $\gamma$-domain (as the slow modes), as expected in the stability analysis in Section~\ref{sec:DT-DOB-Stability}.  

\subsection{Simulation 2: Selection of $\PPRM_{\mathsf n}^\ddSF(z;\Delta)$}

This subsection examines how the discretization method for $\PPRM_{\nnSF}(s)$ affects the stability of the overall system. 
To proceed, we discretize $\CCRM(s)$ using the FDM, and take two types of DT nominal model (from the same CT nominal model): $\PPRM_{{\mathsf n},{\sf BDM}}^\ddSF(z;\Delta)$ and $\PPRM_{{\mathsf n},{\sf BT}}^\ddSF(z;\Delta)$.
Since both DT nominal models are biproper, a 1st-order DT Q-filter \eqref{eq:DT-DOB-Q1st} can be employed for the construction.
It is further noticed that $\PPRM_{{\mathsf n},{\sf BDM}}^\ddSF(z;\Delta)$ satisfies the requirement of the proposed design guideline; in contrast, when $\PPRM_{{\mathsf n},{\sf BT}}^\ddSF(z;\Delta)$ is used, there is ``no'' 1st-order DT Q-filter \eqref{eq:DT-DOB-Q1st} that makes $\Psi_{\sf fast}^\ddSF(z)$ Schur (by Proposition~\ref{prop:Simp}).
In the simulation, the following two 1st-order DT Q-filters are taken into account: $\QQRM_{\sf 1st,LBW}^\ddSF(z;\Delta) = {0.15}/\big({(z-1) + 0.15}\big)$ and $\QQRM_{\sf 1st,SBW}^\ddSF(z;\Delta) = {0.015}/\big({(z-1) + 0.015}\big)$ 
where the former is obtained by the design guideline in Section~\ref{sec:DT-DOB-Design} for $\PPRM_{{\mathsf n},{\sf BDM}}^\ddSF(z;\Delta)$, whereas the latter is selected to have much smaller bandwidth than the former one. 

\begin{figure}[!t]
	\begin{center}
		{
     		\subfigure[DT-DOB with $(\PPRM_{{\mathsf n},{\sf BDM}}^\ddSF,\QQRM^\ddSF_{\sf 1st,LBW})$]	{\includegraphics[width=0.34\textwidth]{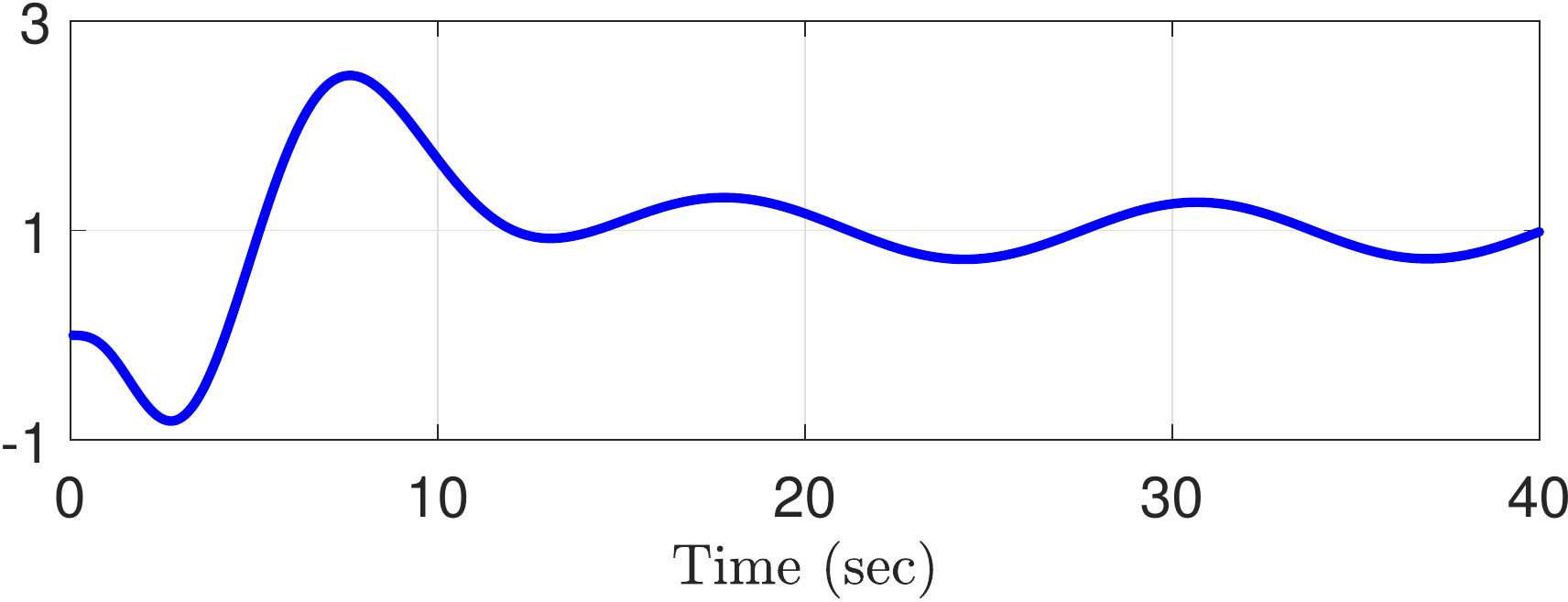}}
			\subfigure[DT-DOB with $(\PPRM_{{\mathsf n},{\sf BT}}^\ddSF,\QQRM^\ddSF_{\sf 1st,LBW})$]	{\includegraphics[width=0.34\textwidth]{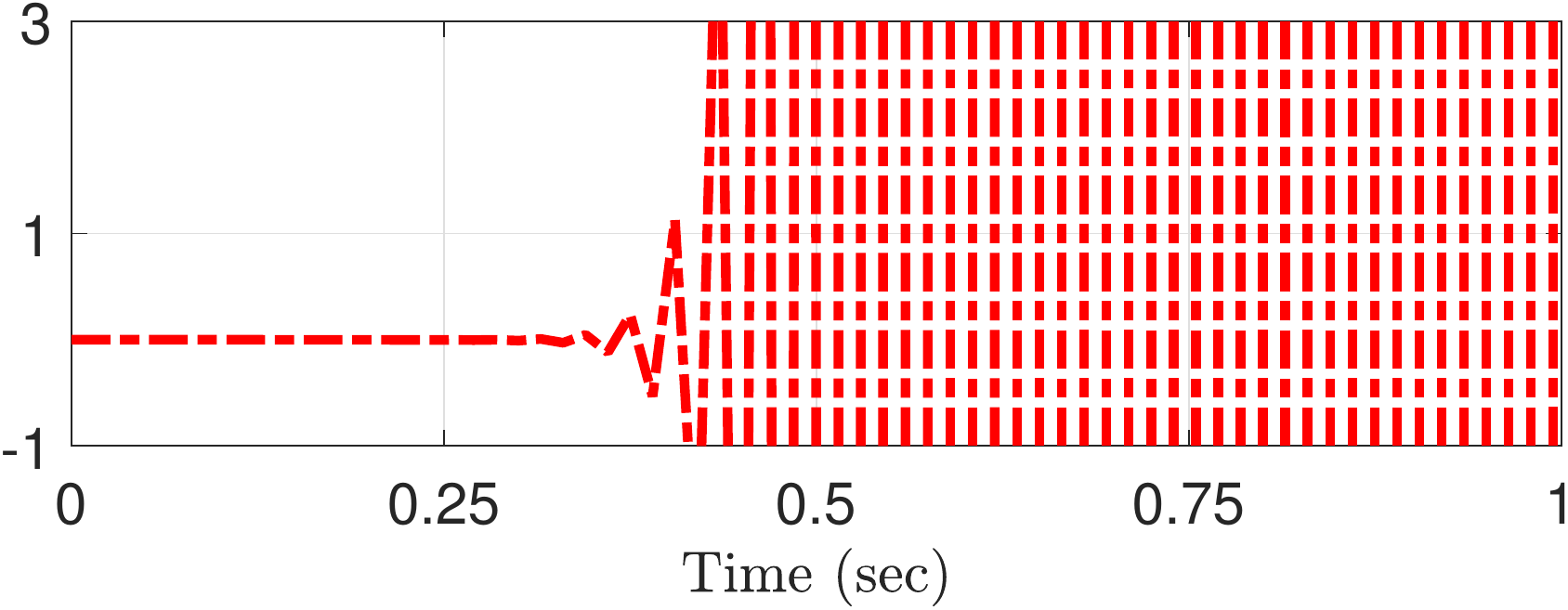}}	\\			
			\subfigure[DT-DOB with $(\PPRM_{{\mathsf n},{\sf BT}}^\ddSF,\QQRM^\ddSF_{\sf 1st,SBW})$]	{\includegraphics[width=0.34\textwidth]{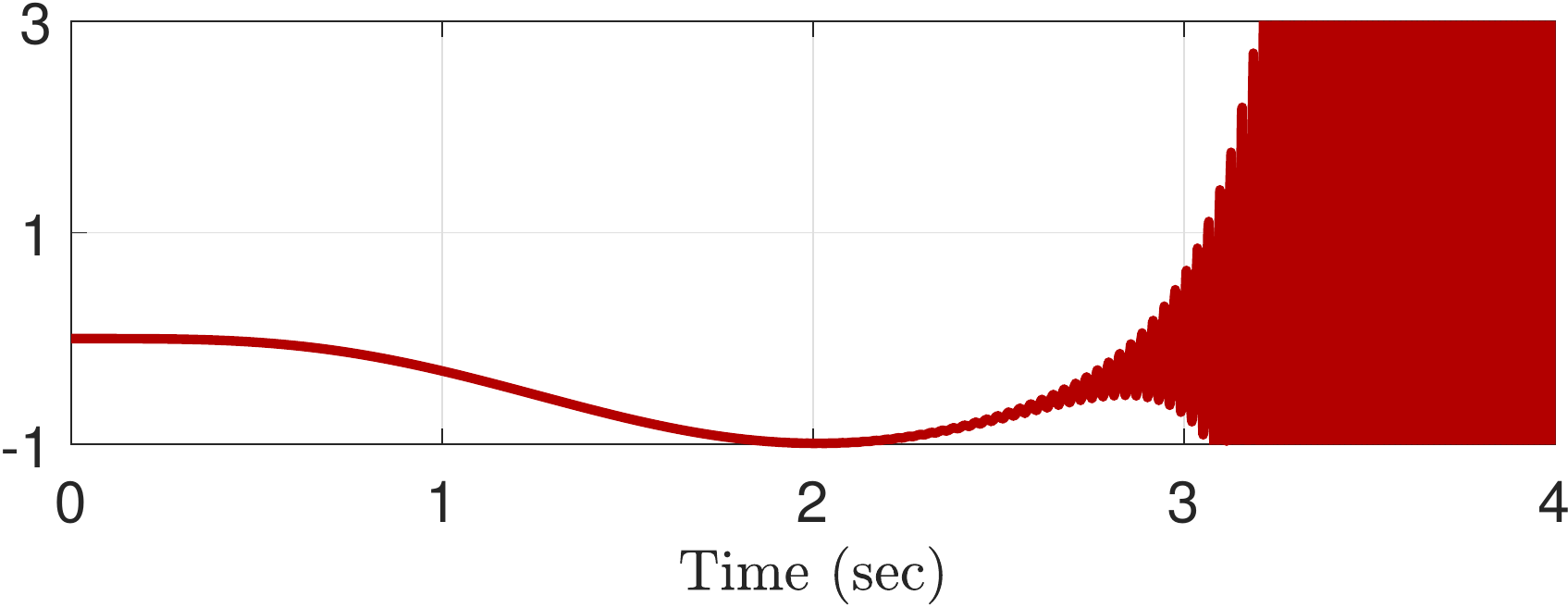}}\\
		}
	\end{center}
	\caption{Step responses $y(t)$ of DT-DOB controlled systems with different discretization methods for DT nominal model}\label{fig:DT-DOB-SIM2}
\end{figure}

We now construct three types of the DT-DOB for comparison;
the first one is constructed following the proposed design guideline, which results in $\PPRM_{{\mathsf n}}^\ddSF(z;\Delta) = \PPRM_{{\mathsf n},{\sf BDM}}^\ddSF(z;\Delta)$ and $\QQRM^\ddSF(z;\Delta) = \QQRM_{\sf 1st, LBW}^\ddSF(z;\Delta)$; on the other hand, the remaining two ones are designed with the same DT nominal model $\PPRM_{{\mathsf n}}^\ddSF(z;\Delta) = \PPRM_{{\mathsf n},{\sf BT}}^\ddSF(z;\Delta)$ but with different DT Q-filters, $\QQRM^\ddSF(z;\Delta) = \QQRM_{\sf 1st,LBW}^\ddSF(z;\Delta)$ and $\QQRM^\ddSF(z;\Delta) =\QQRM_{\sf 1st,SBW}^\ddSF(z;\Delta)$, respectively.  
Fig.~\ref{fig:DT-DOB-SIM2} depicts the time-domain simulations for these DT-DOB controlled systems.
It is seen in the figure that the DT nominal model obtained by the BT yields instability of the closed-loop system  regardless of the bandwidth of the DT Q-filter.
This phenomenon is mainly because Item (c) in Theorem~\ref{thm:DT-DOB-Main} becomes violated  by the use of the BT (see also Proposition~\ref{prop:Simp}).
The simulation result emphasizes the importance of selecting the way of discretization methods for $\PPRM_{{\mathsf n}}^\ddSF(z;\Delta)$ in the DT-DOB design.

\section{Conclusion}

In this paper, we have presented a generalized framework for robust stability analysis of DT-DOB controlled systems. 
Rather than utilizing the small-gain theorem in the analysis, our result is carried out by investigating the limiting behavior of the DT-DOB controlled system as the sampling proceeds fast, based on a general expression of the DT-DOBC.
As a consequence, the stability condition derived here is necessary and sufficient under fast sampling except in a degenerative case and can be applied to a large class of the DT-DOB controlled systems.
It is also revealed in an explicit manner that the stability of the overall system is strongly related with the sampling zeros, the DT Q-filter, the discretization methods for the nominal model, and the model uncertainty.  
In order to satisfy the stability condition against arbitrarily large but bounded parametric uncertainty, a systematic design guideline for the DT-DOBC has been newly presented.

\appendix

\section{Proof of Lemma~\ref{lem:DeltaStar}}
It is readily seen that $\inf\{\bigcup_{\PPRM \in\mathcal{P}}\mathcal{T}_{\PPRM,1}\}>0$, since, under the assumption, the spectral radius of $A$ has a uniform bound independent of $\PPRM \in \mathcal{P}$.
Then the lemma can be proved by showing that there is a positive constant $\underline{\Delta}^\star$ (independent of $\PPRM$) such that $C^\ddSF B^\ddSF({\Delta})$ is nonzero for all $\PPRM\in \mathcal{P}$ and for all $\Delta \in (0,\underline{\Delta}^\star)$ (so that $\inf\{\bigcup_{\PPRM \in {\mathcal{P}}} {\mathcal{T}}_{\PPRM,2}\}>0$). 
Noting that $\int_{0}^\Delta e^{A\tau}\ddRM\tau = \Delta (I + (1/2!)(A\Delta) + (1/3!) (A\Delta)^2 + \cdots)$ and  $CA^{i}B=0$ for all $i=0,\dots,\nu-2$, one has
\begin{align*}
& C^\ddSF B^\ddSF(\Delta) = C \Delta \left( I + \frac{1}{2!} (A\Delta) + \frac{1}{3!} (A \Delta)^2 + \cdots  \right) B\\
& ~~\quad = C \Delta \left(\frac{1}{\nu!}(A\Delta)^{\nu-1} + \frac{1}{(\nu+1)!}(A\Delta)^{\nu} + \cdots \right) B\\
& ~~\quad = \frac{\Delta^{\nu}}{\nu!} \left(  CA^{\nu-1} B + p(\Delta) \right) = \frac{\Delta^\nu}{\nu!}(g + p(\Delta))
\end{align*}
where $p(\Delta):= \Delta \big( (\nu!/(\nu+1)!) C A^{\nu} B + (\nu!/(\nu+2)!) CA^{\nu+1}B \Delta + \cdots \big)$. 
Here note that $\nu !/(\nu+j)! = 1/\big((\nu+1)\times \dots \times (\nu+j)\big) < 1/(j-2)!$ for all $j=1,2,\dots$ (for simplicity, let $(-1)! := 1$).
Thus we have
\begin{align*}
& \| p (\Delta)\|\\
& \leq \Delta \|C A^{\nu}\| \left( \frac{\nu!}{(\nu+1)!}\| I\| + \frac{\nu!}{(\nu+2)!} \|A \| \Delta + \cdots \right) \|B\| \\
& \leq \Delta \|C A^{\nu}\| \left( 1 + \|A \| \Delta + \frac{1}{2!} (\|A \| \Delta)^2 + \cdots \right) \|B\|\\
& = \Delta \|C A^{\nu}\| e^{\|A\| \Delta} \|B\| =: \overline{p}(\Delta).
\end{align*}
It is pointed out that the rightmost term $\overline{p}(\Delta)$ is a continuous function of $\Delta$ that converges to zero as $\Delta\rightarrow 0^+$.
In addition, under Assumption~\ref{asm:DT-DOB-Uncertainty}, the uncertain quantities of $\|C A^\nu\|$, $\|A\|$, and $\|B\|$ are bounded and the bounds can be chosen independent of $\PPRM \in \mathcal{P}$.
From this, we find $\underline{\Delta}^\star>0$ satisfying that $|\overline{p}(\Delta)| < \underline{g}/2$ for all $\Delta \in (0,\underline{\Delta}^\star)$ and for all $\PPRM \in \mathcal{P}$.
This concludes the proof, because for each $\Delta \in (0,\underline{\Delta}^\star)$ and for each $\PPRM \in \mathcal{P}$, $C^\ddSF B^\ddSF(\Delta) = (\Delta^\nu/\nu!) \left(  g + p(\Delta) \right) \geq  (\Delta^\nu/\nu!)(\underline{g} - \|p(\Delta)\|) > (\Delta^\nu/\nu!)(\underline{g}/2)>0$. 


\begin{thebibliography}{34}
\providecommand{\natexlab}[1]{#1}
\providecommand{\url}[1]{\texttt{#1}}
\providecommand{\urlprefix}{URL }
\expandafter\ifx\csname urlstyle\endcsname\relax
  \providecommand{\doi}[1]{doi:\discretionary{}{}{}#1}\else
  \providecommand{\doi}{doi:\discretionary{}{}{}\begingroup
  \urlstyle{rm}\Url}\fi
\providecommand{\eprint}[2][]{\url{#2}}

\bibitem[{{\AA}str{\"o}m et~al.(1984){\AA}str{\"o}m, Hagander, \&
  Sternby}]{AHS84}
{\AA}str{\"o}m, K.~J., Hagander, P., \& Sternby, J. (1984).
\newblock Zeros of sampled systems.
\newblock \emph{Automatica}, \emph{20}(1), 31--38.

\bibitem[{Bertoluzzo et~al.(2004)Bertoluzzo, Buja, \& Stampacchia}]{BBS04}
Bertoluzzo, M., Buja, G.~S., \& Stampacchia, E. (2004).
\newblock Performance analysis of a high-bandwidth torque disturbance
  compensator.
\newblock \emph{{IEEE}/{ASME} Transactions on Mechatronics}, \emph{9}(4),
  653--660.

\bibitem[{Burke et~al.(2006)Burke, Henrion, Lewis, \& Overton}]{BHLO06}
Burke, J.~V., Henrion, D., Lewis, A.~S., \& Overton, M.~L. (2006).
\newblock {HIFOO}--{A} {MATLAB} package for fixed-order controller design and
  $\mathcal{H}_\infty$ optimization.
\newblock In \emph{Proceedings of IFAC Symposium on Robust Control Design}.

\bibitem[{Chen et~al.(2015)Chen, Jiang, \& Tomizuka}]{Chen2015}
Chen, X., Jiang, T., \& Tomizuka, M. (2015).
\newblock Pseudo {Y}oula-{K}ucera parameterization with control of the waterbed
  effect for local loop shaping.
\newblock \emph{Automatica}, \emph{62}, 177--183.

\bibitem[{Chen \& Tomizuka(2010)}]{Chen2010}
Chen, X. \& Tomizuka, M. (2010).
\newblock Optimal plant shaping for high bandwidth disturbance rejection in
  discrete disturbance observers.
\newblock In \emph{Proceedings of American Control Conference}, pp. 2641--2646.

\bibitem[{Chen \& Tomizuka(2012)}]{CT12}
Chen, X. \& Tomizuka, M. (2012).
\newblock A minimum parameter adaptive approach for rejecting multiple
  narrow-band disturbances with application to hard disk drives.
\newblock \emph{IEEE Transactions on Control Systems Technology}, \emph{20}(2),
  408--415.

\bibitem[{Choi et~al.(2016)Choi, Choi, Kong, \& Hyun}]{CCKH16}
Choi, J., Choi, H., Kong, K., \& Hyun, D.~J. (2016).
\newblock An adaptive disturbance observer for precision control of
  time-varying systems.
\newblock In \emph{Proceedings of IFAC Symposium on Mechatronics Systems}, pp.
  240--245.

\bibitem[{Choi et~al.(2003)Choi, Yang, Chung, Kim, \& Suh}]{CYCKS03}
Choi, Y., Yang, K., Chung, W.~K., Kim, H.~R., \& Suh, I.~H. (2003).
\newblock On the robustness and performance of disturbance observers for
  second-order systems.
\newblock \emph{{IEEE} Transactions on Automatic Control}, \emph{48}(2),
  315--320.

\bibitem[{Flanigan(1983)}]{Flanigan83}
Flanigan, F.~J. (1983).
\newblock \emph{Complex {V}ariables}.
\newblock Dover Publication.

\bibitem[{Franklin et~al.(1998)Franklin, Powell, \& Workman}]{FPW98}
Franklin, G.~F., Powell, J.~D., \& Workman, M.~L. (1998).
\newblock \emph{Digital {C}ontrol of {D}ynamic {S}ystems}.
\newblock Addison-{W}esley.

\bibitem[{Godler et~al.(2002)Godler, Honda, \& Ohnishi}]{GHO02}
Godler, I., Honda, H., \& Ohnishi, K. (2002).
\newblock Design guidelines for disturbance observer's filter in discrete time.
\newblock In \emph{Proceedings of International Workshop on Advanced Motion
  Control}, pp. 390--395.

\bibitem[{Kempf \& Kobayashi(1999)}]{KK99}
Kempf, C.~J. \& Kobayashi, S. (1999).
\newblock Disturbance observer and feedforward design for a high-speed
  direct-drive positioning table.
\newblock \emph{{IEEE} Transactions on Control Systems Technology},
  \emph{7}(5), 513--526.

\bibitem[{Kokotovic et~al.(1999)Kokotovic, Khalil, \& O'reilly}]{KKO99}
Kokotovic, P., Khalil, H.~K., \& O'reilly, J. (1999).
\newblock \emph{Singular perturbation methods in control: {A}nalysis and
  design}.
\newblock Society for Industrial and Applied Mathematics.

\bibitem[{Kong \& Tomizuka(2013)}]{Kong2013}
Kong, K. \& Tomizuka, M. (2013).
\newblock Nominal model manipulation for enhancement of stability robustness
  for disturbance observer-based control systems.
\newblock \emph{International Journal of Control, Automation and Systems},
  \emph{11}(1), 12--20.

\bibitem[{Kwon \& Chung(2003)}]{KC03b}
Kwon, S. \& Chung, W.~K. (2003).
\newblock A discrete-time design and analysis of perturbation observer for
  motion control applications.
\newblock \emph{{IEEE} Transactions on Control Systems Technology},
  \emph{11}(3), 399--407.

\bibitem[{Lee et~al.(2012)Lee, Joo, \& Shim}]{LJS12}
Lee, C., Joo, Y., \& Shim, H. (2012).
\newblock Analysis of discrete-time disturbance observer and a new {Q}-filter
  design using delay function.
\newblock In \emph{Proceedings of International Conference on Control,
  Automation and Systems}, pp. 556--561.

\bibitem[{Lee \& Tomizuka(1996)}]{Lee1996}
Lee, H.~S. \& Tomizuka, M. (1996).
\newblock Robust motion controller design for high-accuracy positioning
  systems.
\newblock \emph{{IEEE} Transactions on Industrial Electronics}, \emph{43}(1),
  48--55.

\bibitem[{Li et~al.(2014)Li, Yang, Chen, \& Chen}]{LYCC14}
Li, S., Yang, J., Chen, W.~H., \& Chen, X. (2014).
\newblock \emph{{Disturbance Observer-based Control: Methods and
  Applications}}.
\newblock CRC Press.

\bibitem[{Litkouhi \& Khalil(1985)}]{Litkouhi1985}
Litkouhi, B. \& Khalil, H.~K. (1985).
\newblock Multirate and composite control of two-time-scale discrete-time
  systems.
\newblock \emph{{IEEE} Transactions on Automatic Control}, \emph{30}(7),
  645--651.

\bibitem[{Ohnishi(1987)}]{Ohnishi87}
Ohnishi, K. (1987).
\newblock A new servo method in mechatronics.
\newblock \emph{Transactions on Japan Society of Electrical Engineers (in
  {J}apanese)}, \emph{107-D}, 83--86.

\bibitem[{Park et~al.(2015)Park, Joo, Lee, \& Shim}]{Park2015}
Park, G., Joo, Y., Lee, C., \& Shim, H. (2015).
\newblock On robust stability of disturbance observer for sampled-data systems
  under fast sampling: {A}n almost necessary and sufficient condition.
\newblock In \emph{Proceedings of {IEEE} Conference on Decision and Control},
  pp. 7536--7541.

\bibitem[{Park \& Shim(2015)}]{Park2015a}
Park, G. \& Shim, H. (2015).
\newblock A generalized framework for robust stability analysis of
  discrete-time disturbance observer for sampled-data systems: {A} fast
  sampling approach.
\newblock In \emph{Proceedings of International Conference on Control,
  Automation and Systems}, pp. 295--300.

\bibitem[{Phillips \& Nagle(2007)}]{PN07}
Phillips, C.~L. \& Nagle, H.~T. (2007).
\newblock \emph{Digital {C}ontrol {S}ystem {A}nalysis and {D}esign}.
\newblock Prentice-Hall.

\bibitem[{Sariyildiz \& Ohnishi(2015)}]{SO2015}
Sariyildiz, E. \& Ohnishi, K. (2015).
\newblock Stability and robustness of disturbance-observer-based motion control
  systems.
\newblock \emph{{IEEE} Transactions on Industrial Electronics}, \emph{62}(1),
  414--422.

\bibitem[{Shim \& Jo(2009)}]{SJ09}
Shim, H. \& Jo, N.~H. (2009).
\newblock An almost necessary and sufficient condition for robust stability of
  closed-loop systems with disturbance observer.
\newblock \emph{Automatica}, \emph{45}(1), 296--299.

\bibitem[{Shim \& Joo(2007)}]{SJ07}
Shim, H. \& Joo, Y. (2007).
\newblock State space analysis of disturbance observer and a robust stability
  condition.
\newblock In \emph{Proceedings of IEEE Conference on Decision and Control}, pp.
  2193--2198.

\bibitem[{Shim et~al.(2016)Shim, Park, Joo, Back, \& Jo}]{SPJBJ16}
Shim, H., Park, G., Joo, Y., Back, J., \& Jo, N.~H. (2016).
\newblock Yet another tutorial of disturbance observer: {R}obust stabilization
  and recovery of nominal performance.
\newblock \emph{Control Theory Technology}, \emph{14}(3), 237--249.
\newblock ({S}pecial issue on disturbance rejection: {A} snapshot, a necessity,
  and a beginning), arXiv preprint arXiv:1601.02075.

\bibitem[{Tesfaye et~al.(2000)Tesfaye, Lee, \& Tomizuka}]{TLT00}
Tesfaye, A., Lee, H.~S., \& Tomizuka, M. (2000).
\newblock A sensitivity optimization approach to design of a disturbance
  observer in digital motion control systems.
\newblock \emph{{IEEE}/{ASME} Transactions on Mechatronics}, \emph{5}(1),
  32--38.

\bibitem[{Uzunovic et~al.(2018)Uzunovic, Sariyildiz, \&
  Sabanovic}]{Uzunovic2018}
Uzunovic, T., Sariyildiz, E., \& Sabanovic, A. (2018).
\newblock A discussion on discrete implementation of disturbance-observer-based
  control.
\newblock In \emph{{IEEE} International Workshop on Advanced Motion Control},
  pp. 613--618.

\bibitem[{Wie \& Bernstein(1992)}]{WB92}
Wie, B. \& Bernstein, D.~S. (1992).
\newblock Benchmark problems for robust control design.
\newblock \emph{Journal of Guidance, Control, and Dynamics}, \emph{15}(5),
  1057--1059.

\bibitem[{Yang et~al.(2003)Yang, Choi, \& Chung}]{YCC03}
Yang, K., Choi, Y., \& Chung, W.~K. (2003).
\newblock Performance analysis of discrete-time disturbance observer for
  second-order systems.
\newblock In \emph{Proceedings of {IEEE} Conference on Decision and Control},
  pp. 4877--4882.

\bibitem[{Yun et~al.(2016)Yun, Park, Shim, \& Chang}]{Yun2016}
Yun, H., Park, G., Shim, H., \& Chang, H.~J. (2016).
\newblock State-space analysis of discrete-time disturbance observer for
  sampled-data control systems.
\newblock In \emph{Proceedings of American Control Conference}, pp. 4233--4238.

\bibitem[{Yun et~al.(2017)Yun, Shim, \& Chang}]{Yun2017}
Yun, H., Shim, H., \& Chang, H.~J. (2017).
\newblock Singular perturbation for sampled-data systems with fast subsystems.
\newblock In \emph{Proceedings of IFAC World Congress}, volume~50, pp.
  8145--8150.

\bibitem[{Yuz \& Goodwin(2014)}]{YG14}
Yuz, J.~I. \& Goodwin, G.~C. (2014).
\newblock \emph{Sampled-data Models for Linear and Nonlinear Systems}.
\newblock Springer.

\end{thebibliography}
\end{document}